\documentclass[twocolumn, tighten,preprint]{aastex62}
\usepackage{apjfonts}
\usepackage{color}
\usepackage{amsmath, mathrsfs}

\DeclareMathAlphabet{\mathbi}{OT1}{ptm}{bx}{it}
\SetMathAlphabet\mathbi{bold}{OT1}{ptm}{bx}{it}

\begin{document}

\title{\bf Untangling Optical Emissions of the Jet and Accretion Disk in the Flat-Spectrum Radio Quasar 3C 273 with Reverberation Mapping Data}

\author[0000-0001-5841-9179]{Yan-Rong Li}
\affiliation{Key Laboratory for Particle Astrophysics, Institute of High 
Energy Physics, Chinese Academy of Sciences, \\19B Yuquan Road, 
Beijing 100049, China; \href{mailto:liyanrong@mail.ihep.ac.cn}{liyanrong@mail.ihep.ac.cn}, \href{mailto:wangjm@mail.ihep.ac.cn}{wangjm@mail.ihep.ac.cn}}

\author{Zhi-Xiang Zhang}
\affiliation{Department of Astronomy, Xiamen Univeristy, Xiamen, Fujian 361005, China}

\author[0000-0002-2006-1615]{Chichuan Jin}
\affiliation{National Astronomical Observatories of China, Chinese 
Academy of Sciences, A20 Datun Road, Beijing 100012, China}

\author[0000-0002-5830-3544]{Pu Du}
\affiliation{Key Laboratory for Particle Astrophysics, Institute of High 
Energy Physics, Chinese Academy of Sciences, \\19B Yuquan Road, 
Beijing 100049, China; \href{mailto:liyanrong@mail.ihep.ac.cn}{liyanrong@mail.ihep.ac.cn}, \href{mailto:wangjm@mail.ihep.ac.cn}{wangjm@mail.ihep.ac.cn}}

\author{Lang Cui}
\author{Xiang Liu}
\affiliation{Xinjiang Astronomical Observatory, Chinese Academy of Sciences, 150 Science 1-Street, Urumqi 830011, China}

\author[0000-0001-7617-4232]{Jian-Min Wang}
\affiliation{Key Laboratory for Particle Astrophysics, Institute of High 
Energy Physics, Chinese Academy of Sciences, \\19B Yuquan Road, 
Beijing 100049, China; \href{mailto:liyanrong@mail.ihep.ac.cn}{liyanrong@mail.ihep.ac.cn}, \href{mailto:wangjm@mail.ihep.ac.cn}{wangjm@mail.ihep.ac.cn}}
\affiliation{National Astronomical Observatories of China, Chinese 
Academy of Sciences, A20 Datun Road, Beijing 100012, China}
\affiliation{School of Astronomy and Space Science, University of Chinese Academy of Sciences, 
19A Yuquan Road, Beijing 100049, China}

\begin{abstract}
3C 273 is an intensively monitored flat-spectrum radio quasar with both a beamed jet and blue bump together with broad emission lines.
The coexistence of the comparably prominent jet and accretion disk leads to complicated variability properties.
Recent reverberation mapping monitoring for 3C~273 revealed that the optical continuum shows a distinct long-term trend that does not have 
a corresponding echo in the H$\beta$ fluxes. We compile multi-wavelength monitoring data from the {\it Swift} archive and other ground-based programs 
and clearly find two components of emissions at optical wavelength. One component stems from the accretion disk itself and the other 
component can be ascribed to the jet contribution, which also naturally accounts for the non-echoed trend in reverberation mapping data. 
We develop an approach to decouple the optical emissions from the jet and accretion disk in 3C~273
with the aid of multi-wavelength monitoring data. By assuming the disk emission has a negligible polarization 
in consideration of the low inclination of the jet, the results show that the jet contributes a fraction of $\sim$10\% at the minimum 
and up to $\sim$40\% at the maximum to the total optical emissions.
This is the first time to provide a physical interpretation to the ``detrending'' manipulation 
conventionally adopted in reverberation mapping analysis. Our work also illustrates the importance of appropriately analyzing 
variability properties in cases of coexisting jets and accretion disks.  
\end{abstract}

\section{Introduction}
3C 273 is an iconic object in extragalactic astronomy because of its historic role in discovering the first quasar and the first extragalactic radio jet 
(\citealt{Hazard2018}). It is classified to be a flat-spectrum radio quasar that has both a prominent blue bump together with broad emission lines, 
indicative of an accretion disk radiating at its nucleus (\citealt{Paltani1998, Kriss1999, Turler1999, Soldi2008}), and a beamed jet, a characteristic typical for 
blazar objects  (\citealt{Davis1991, Bahcall1995, Abraham1999, Perley2017}).
However, unlike blazar objects, the optical polarization of 3C 273 is distinctively at low level (in average $p<$1\%; 
\citealt{Stockman1984, Berriman1990, Brindle1990, Marin2014, Hutsemekers2018}).
These lines of observations make 3C 273 an archetype of active galactic nuclei (AGNs) in general and a good laboratory to study various 
(if not all) AGN phenomena in particular. With addition of its large brightness ($K$-band magnitude $\sim10$) and mild proximity ($z=0.158$),  
3C 273 had been intensively monitored and studied across almost all the wavelength bands 
(e.g., see \citealt{Courvoisier1998} for a review).

\begin{deluxetable*}{ccccccccc}
\tablecolumns{9}
\tabletypesize{\footnotesize}
\tablecaption{Monitoring Data of 3C 273. \label{tab_data}}
\tablehead{
 \colhead{Source}   & 
 \colhead{Filter} &
 \colhead{Wavelength} &
 \multicolumn{2}{c}{Observation Period} &
 \colhead{~~~~$N_{\rm obs}$~~~~} &
 \colhead{~~~~$\Delta t_{\rm ave}$~~~~}  & 
 \colhead{~~~~$\Delta t_{\rm med}$~~~~}    &
 \colhead{~~~Ref~~~}\\\cline{4-5} 
 \colhead{}  &
 \colhead{}  &
 \colhead{}  &
 \colhead{JD-2,450,000} &
 \colhead{Date} &
 \colhead{}  &
 \colhead{(day)} &
 \colhead{(day)}
}
\startdata
\it Swift     &      UVW2    & 1928~\AA & 3562.152$-$8553.836 & 2005 Jul$-$2019 Mar & 246          &  20.3    & 2.9  & \nodata\\
\it Swift     &      UVM2    & 2246~\AA & 3562.152$-$8582.797 & 2005 Jul$-$2019 Apr & 232          &  21.6    & 2.9  &\nodata\\
\it Swift     &      UVW1    & 2600~\AA & 3562.152$-$8555.820 & 2005 Jul$-$2019 Mar & 225          &  22.2    & 3.1  &\nodata\\
\it Swift     &  \it U       & 3465~\AA & 3562.086$-$8580.789 & 2005 Jul$-$2019 Apr & 194          &  25.9    & 1.9  &\nodata\\
\it Swift     &  \it B Swift       & 4392~\AA & 3562.086$-$8304.324 & 2005 Jul$-$2018 Jul &  111   &  42.7   &  1.1 &\nodata\\
SMARTS        &  \it B       & 4450~\AA & 4677.497$-$7856.656 & 2008 Jul$-$2017 Apr & 363          &  8.8    &  2.9 & 3        \\
RM\tablenotemark{$a$}            & \nodata    & 5100~\AA & 4795.018$-$8305.687 & 2008 Nov$-$2018 Jul & 285 &   11.4   &  2.1 & 4 \\
\it Swift     &  \it V~Swift       & 5468~\AA & 3562.086$-$8304.324 & 2005 Jul$-$2018 Jul & 225            &   21.1   &  1.9 &\nodata\\
ASAS-SN   &  $V$       & 5510~\AA & 5956.146$-$8449.136 & 2012 Jan$-$2018 Nov & 988 &   2.5   & 0.002  & 2      \\
SMARTS    &  $V$       & 5510~\AA & 4677.499$-$7856.658 & 2008 Jul$-$2017 Apr & 365 &   4.7   & 2.1  & 3            \\
RM        &  $V$       & 5510~\AA & 4795.020$-$8305.697 & 2008 Nov$-$2018 Jul & 306 &   11.9   & 2.0  & 4 \\
SMARTS    &  $R$       & 6580~\AA & 4537.596$-$7856.660 & 2008 Jul$-$2017 Apr & 371 &   8.9   & 2.9  & 3            \\
SMARTS    &  $J$       & 12200~\AA & 4501.786$-$7091.743 & 2008 Feb$-$2015 Mar & 306 &  8.5    & 2.1  & 3            \\
OVRO      & \nodata    & 200~$\mu$m & 4473.983$-$8874.077  &  2008 Jan$-$2020 Jan&760 & 5.8 &  3.0 & 5
\enddata
\tablerefs{ (1) \cite{Drake2009}; (2) \cite{Shappee2014} and \cite{Kochanek2017}; (3) \cite{Bonning2012}; 
            (4) \cite{Zhang2019}; (5) \cite{Richards2011}.
            }
\tablenotetext{a}{``RM'' means that the data is from a reverberation mapping campaign presented in \cite{Zhang2019}.}
\end{deluxetable*}

The coexistence of the both comparably prominent jet and accretion disk results in the complicated emergent spectrum and variability properties
(e.g., \citealt{Stevens1998, Grandi2004,Turler2006, Soldi2008, Chidiac2016, Plavin2019}). 
By analyzing the monitoring data of the optical polarization, \cite{Impey1989} suggested that there is a miniblazar in 3C 273 that contributes about 10\%
of the optical flux densities. It is also this miniblazar component (with a high polarization) diluted by the disk emissions, leading to 
the low-level, highly variable polarization observed in 3C~273. For the blue bump of 3C 273, \cite{Paltani1998} proposed a decomposition 
of blue and red components. The former was explained by the thermal accretion disk emission and the latter was ascribed to the jet origin.
Based on spectral fitting, \cite{Grandi2004} similarly decomposed
the X-ray spectrum of 3C 273 into two major contributions: a thermal component arising from the accretion disk and hot corona, and 
a non-thermal component arising from the jet. 

Recently, \cite{Zhang2019} presented an optical reverberation mapping campaign
for 3C 273 and found that the optical continuum shows a distinct long-term trend that does not have a corresponding echo in the 
light curves of the broad emission lines (including H$\beta$, H$\gamma$, and \ion{Fe}{2}; see Figure~4 therein). 
This is in contradiction to the well observationally tested reverberation mapping scenario
that broad emission lines stem from gaseous regions photoionized by the ionizing continuum and thereby the variations of broad 
emission lines closely follow these of the continuum (e.g., \citealt{Peterson1993}).
To alleviate biases in reverberation mapping analysis, \cite{Zhang2019} employed a linear polynomial to fit 
this long-term trend and artificially subtracted the best linear fit from the original light curve of the 
optical continuum (see also \citealt{Wang2020}). Such a ``detrending'' procedure was conventionally manipulated in the presence of non-echoed trends
in reverberation mapping observations (\citealt{Welsh1999, Denney2010, Li2013, Peterson2014}). However, to our best knowledge, 
there is not yet a satisfactory physical interpretation for this ``detrending'' operation.

Inspired by the above investigations, in this paper we link the non-echoed trend in 3C 273 with the jet contaminations 
at optical wavelength. The wealth of monitoring data and reverberation mapping observations for 3C~273 allows us to test for this possibility 
in a solid foundation. This is also a first attempt to give a physical explanation for the ``detrending'' operation in 
reverberation mapping analysis.

The paper is organized as follows. Section 2 collects publicly available monitoring data of 3C 273. Section 3 discusses several lines of
evidence for the jet contaminations to the optical emissions. In Section 4, we develop a Bayesian decomposition framework and present the 
obtained results for decoupling the jet and disk emissions at optical wavelength. The discussions and conclusions are given in 
Sections 5 and 6, respectively.
For the sake of brevity, when referring to the Julian Date, only the four least significant digits are retained.

\begin{figure*}
\centering 
\includegraphics[width=0.95\textwidth]{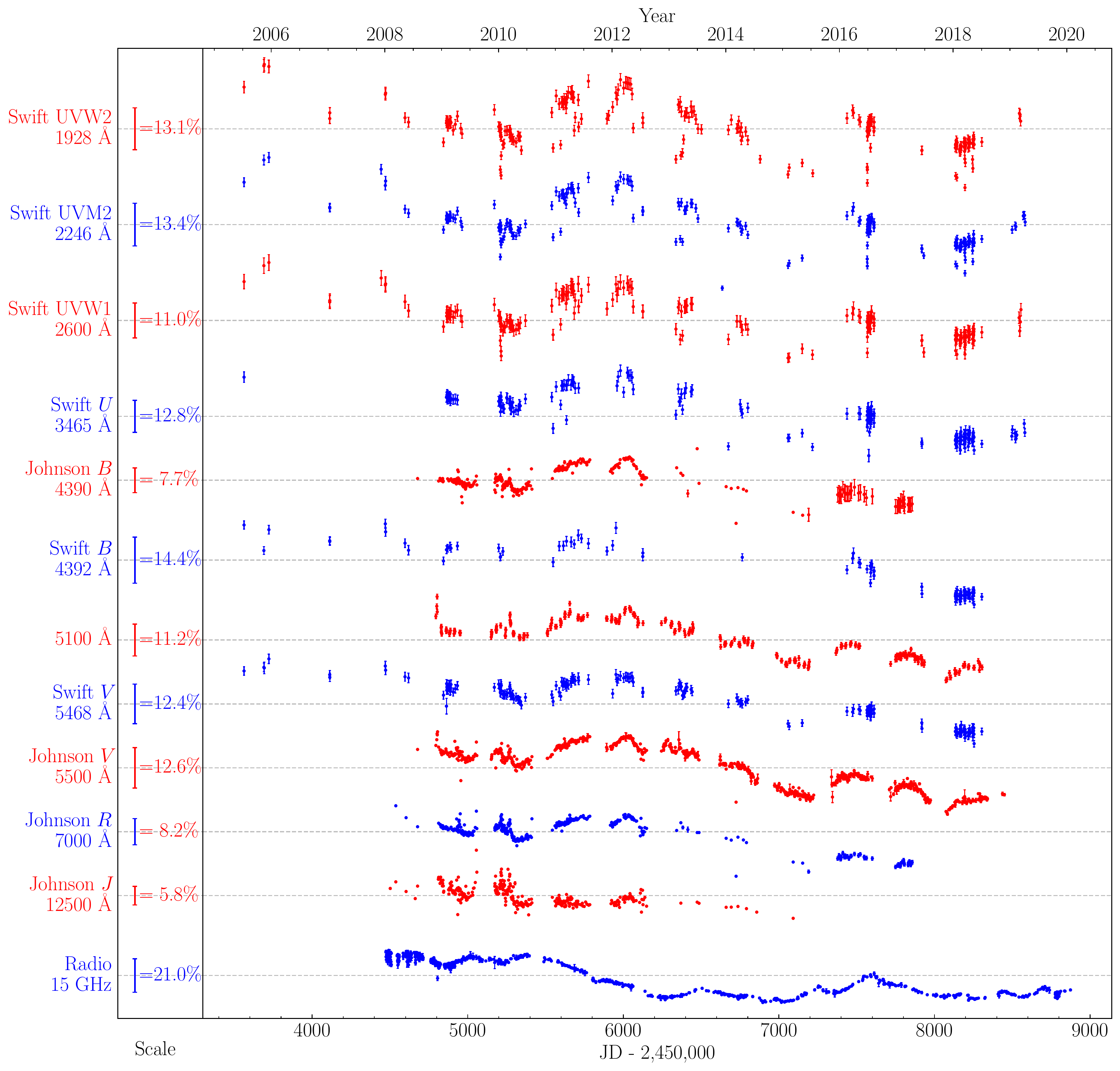}
\caption{Compiled UV, optical, and radio continuum light curves of 3C 273. The scales along the vertical axis show the fractional variations of the 
corresponding light curves.}
\label{fig_lc_all}
\end{figure*}

\section{Data Compilation}
In this section, we collect monitoring data of 3C 273 from various telescopes and sources. All the data compiled here are publicly accessible and most of them 
are directly usable except for the spectroscopic data and the archive UVOT data from the {\it Swift} telescope, which need extra reduction. 
For photometric data with the same filters, an intercalibration is required to account for different apertures adopted in different data sources.
In Figure~\ref{fig_lc_all}, we show all the compiled UV, optical, and radio continuum light curves.

\subsection{Reverberation Mapping Data}
\cite{Zhang2019} reported a reverberation mapping campaign for 3C 273 that synthesized the spectroscopic and photometric data 
from the Steward Observatory spectropolarimetric monitoring project\footnote{The website is at \url{http://james.as.arizona.edu/~psmith/Fermi}.} (\citealt{Smith2009}) and the 
super-Eddington accreting massive black hole (SEAMBH) program (\citealt{Du2014}). The Steward Observatory project utilizes the 2.3 m Bok Telescope on
Kitt Peak and the 1.54 m Kuiper Telescope on Mt. Bigelow in
Arizona. The SEAMBH program utilizes the 2.4 m telescope at the Lijiang Station of the Yunnan
Observatories, Chinese Academy of Sciences.

\begin{figure}
\centering 
\includegraphics[width=0.48\textwidth]{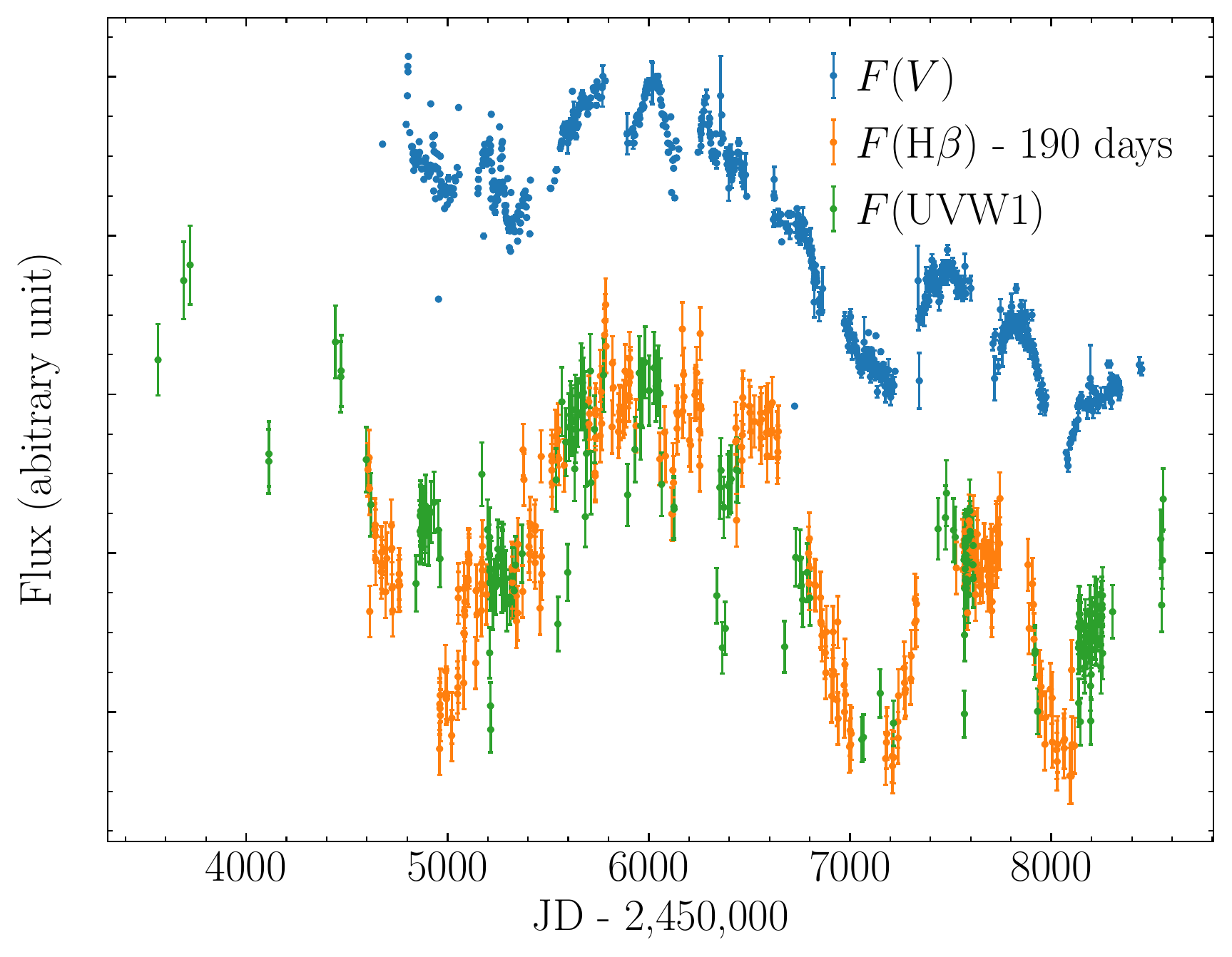}
\caption{A comparison between the light curves of the H$\beta$ fluxes, {\it Swift} UVW1 band photometry, and Johnson $V$-band photometry. 
All the light curves are scaled and shifted for clarity and the H$\beta$ light curve 
is additionally shifted backward by a time of 190 days.}
\label{fig_comp}
\end{figure}

The campaign spanned from November 2008 to March 2018 and took a total of 296 epochs of observations. We directly use the light curves 
of the $V$-band photometry, 5100~{\AA} flux densities, and H$\beta$ fluxes  reduced by \cite{Zhang2019}. As can be seen clearly in 
Figures~\ref{fig_lc_all} and \ref{fig_comp}, the light curves of the $V$-band photometry and 5100~{\AA} flux densities show a long-term 
declining trend whereas the light curve of the H$\beta$ fluxes does not show such a trend, meaning that the H$\beta$ emission-line region 
does not reverberate to this long-term trend.

\begin{figure*}
\centering 
\includegraphics[width=1.0\textwidth]{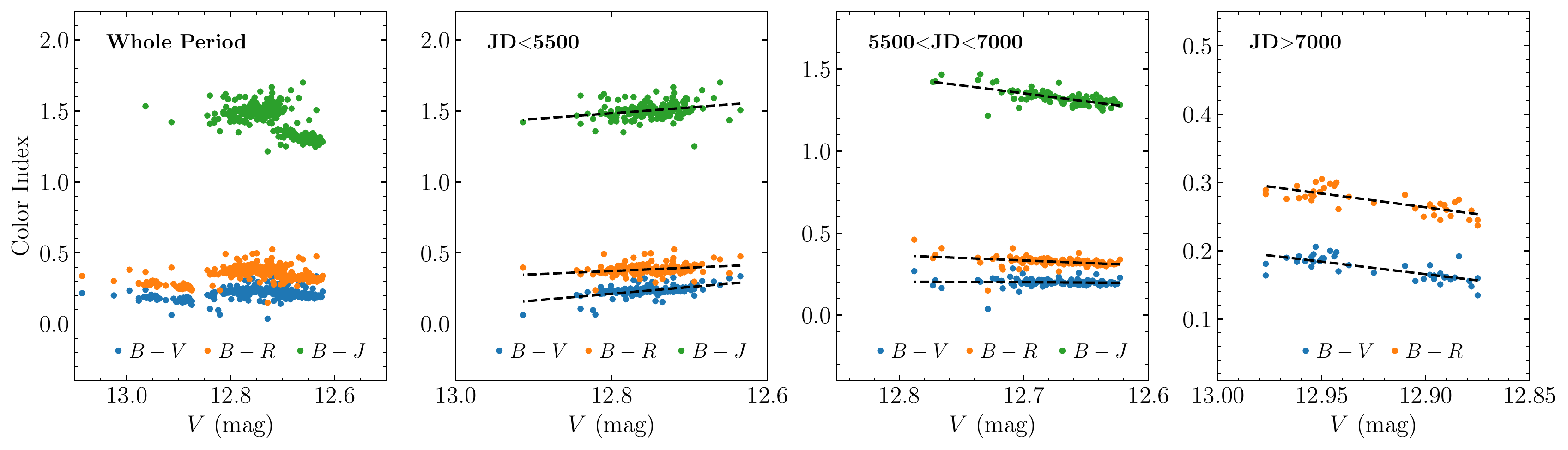}
\caption{Color variations of 3C 273 with $V$-band magnitude using SMARTS data for the periods of the whole, JD$<$5500, 5500$<$JD$<$7000, 
and JD$>$7000 from the left to right panels, respectively. Dashed lines represent simple linear fits. In the rightmost panel, there are 
no $J$-band measurement points.}
\label{fig_color}
\end{figure*}

\subsection{Optical/UV Photometric Data}
Besides the $V$-band photometric data from the reverberation mapping campaign in \cite{Zhang2019}, there are also other archival databases, monitoring programs,
and time-domain surveys that cover 3C 273.

\begin{itemize}
\item The Small and Moderate Aperture Research Telescope System (SMARTS) monitoring program\footnote{The website is at \url{http://www.astro.yale.edu/smarts/glast/home.php}.} 
(\citealt{Bonning2012}). The program was conducted with 
the 1.3 m telescope at the Cerro Tololo Interamerican Observatory, which took photometry at five wavelength bands ($B$, $V$, $R$, $J$, $K$) simultaneously. 
This allow us to study the optical color variations of 3C 273. The $K$-band data has a relatively sparser cadence and we thus only use the other four 
band data.

\item The All-Sky Automated Survey for Supernovae\footnote{The website is at \url{http://www.astronomy.ohio-state.edu/~assassin/index.shtml}.} (ASAS-SN; \citealt{Shappee2014, Kochanek2017}). 
The ASAS-SN started to monitor 3C 273 at $V$-band since January 2012 and provides a real-time interface to access the $V$-band photometry. Typically, there 
are multiple exposures within one night and we combine those multiple exposures into one measurement.


\item The {\it Swift} archive.  The raw image data of six UVOT filters are open-access, covering the UV/optical from 1928 to 5468~\AA.
We reduced those raw data and measured the photometric fluxes (see Appendix~\ref{sec_swift} for the details of data reduction).  We excluded those apparently 
problematic points which were considered to be caused by the contamination of dust and/or other debris within the instrument (\citealt{Edelson2015}) and 
finally obtained about 230 
epochs of measurements that span from July 2005 to March 2019 for each filters.

\end{itemize}

To convert magnitudes to flux densities, we adopt the zero-magnitude points for $B$, $V$, $R$, and $J$ bands determined by \cite{Johnson1966} as follows: 
$F(B=0)=7.20$, $F(V=0)=3.92$, $F(R=0)=1.76$, and $F(J=0)=0.34$, all with a unit of $10^{-9}{\rm~erg~s^{-1}~cm^{-2}~\text{\AA}^{-1}}$.
In addition, we need to intercalibrate the photometry at Johnson $V$-band from different sources. To this end, we first select the SMARTS $V$-band photometry
as the reference set and then apply a scale factor ($\varphi$)
and flux adjustment ($G$) to the flux densities of the other sources as (e.g., \citealt{Peterson1995, Li2014})
\begin{equation}
F(V) = \varphi F(V)_{\rm obs} + G.
\end{equation}
As such, the light curves from all the sources are aligned into a common scale.
Here, the values of $\varphi$ and $G$ are determined by comparing the closely spaced measurements within 5 days from two data sources.
Note that we do not align the {\it Swift} $V$-band photometry with the other Johnson $V$-band photometry. The intercalibrated $V$-band 
light curve is also shown in Figure~\ref{fig_lc_all}.

\subsection{Radio Data}
We use the radio data from the large-scale, fast-cadence 15 GHz monitoring program with the 40 m telescope at the Owens Valley Radio Observatory (\citealt{Richards2011}),
which began in late 2007 and had a nearly daily cadence (but with seasonal gaps). The program is still ongoing and the latest released data was to 
Jan 28, 2020. There are 
in total 760 epochs of observations by the time of writing.

In Table~\ref{tab_data}, we summarize the basic properties of all the compiled light curve data.

\section{Evidence for Jet Contaminations}
Using the monitoring data collected above, in this section we present four pieces of evidence that the jet emissions at optical wavelengths are non-negligible.

\subsection{UV and Optical Variations}
\label{sec_uv}
In Figure~\ref{fig_comp}, we compare the light curve of the H$\beta$ fluxes with these at the $V$-band and {\it Swift} UVW1 band. 
Previous reverberation mapping observations for H$\beta$ lines had well established that H$\beta$ lines respond to (UV) continuum 
variations with a time delay due to the light traveling time from the central continuum sources to the broad-line regions (\citealt{Kaspi2000,Bentz2013, Du2016}).
The H$\beta$ light curve is therefore scaled and shifted in both flux and time to align with the UV light curve. 
We can find that the variation patterns generally match with each other.
This is consistent with the simple photoionization theory that broad emission lines are reprocessed emissions from the gaseous regions photoionized by the central 
UV/X-ray ionizing  photons. Therefore, emission line variations are just blurred echoes of the ionizing continuum variations with time delays arising from 
light traveling times between the ionizing source and gaseous regions.

On the contrary, the $V$-band light curve shows a distinct long-term declining trend that is absent in the H$\beta$ and UVW1 light curves.
Previous multi-wavelength continuum monitoring of AGNs had clearly demonstrated that AGN variations are tightly correlated throughout UV and optical bands
(e.g., \citealt{Edelson2019}). Such a distinct variation trend in 3C~273 strongly suggests that in addition to the accretion disk emission, 
there has to be another independent component that contributes to the optical emissions.

\begin{figure}
\centering 
\includegraphics[width=0.45\textwidth]{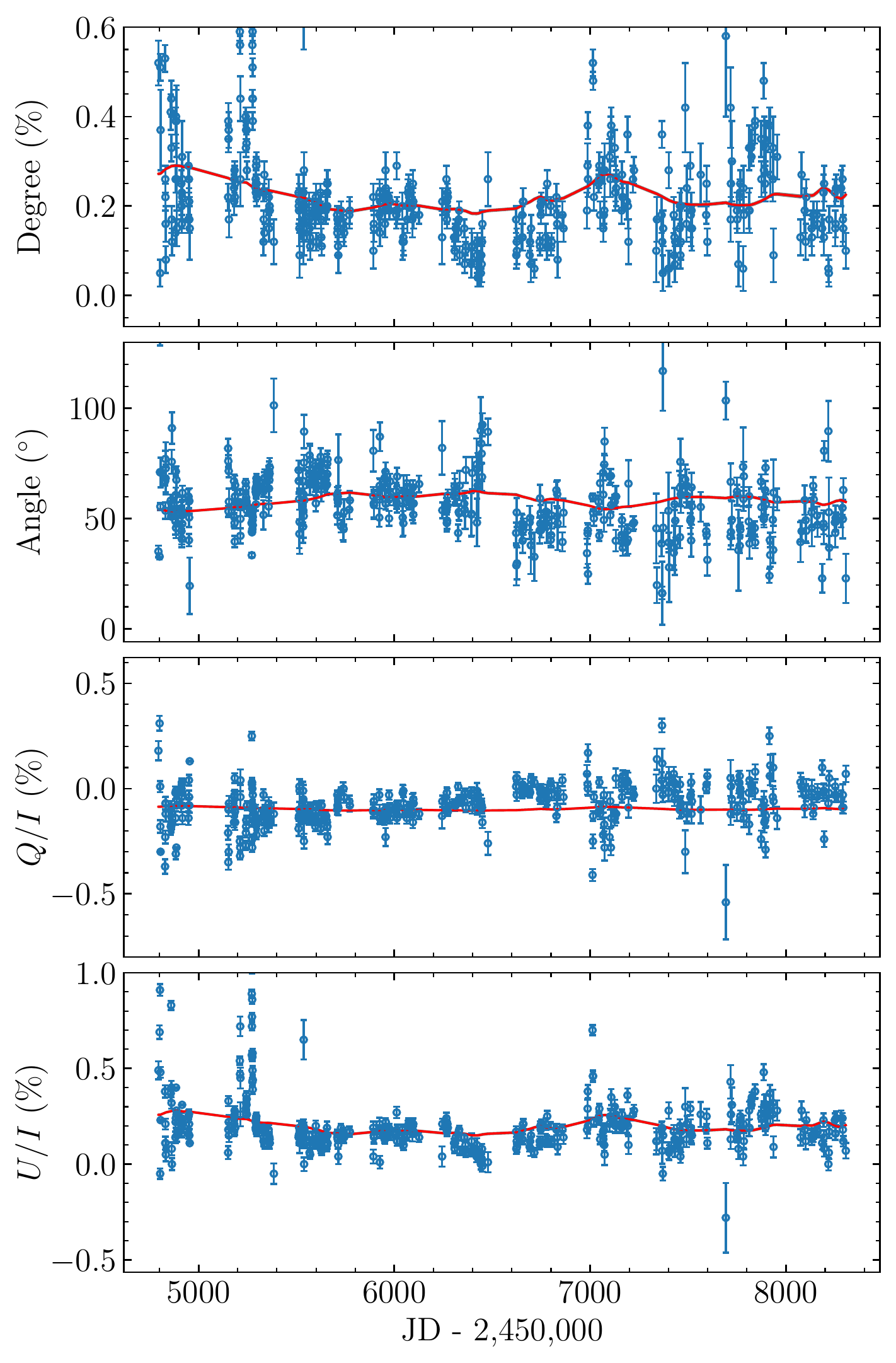}
\caption{Variability of the optical integrated polarizations of 3C~273 measured by the Steward Observatory spectropolarimetric 
monitoring project (\citealt{Smith2009}).
From top to bottom panels are  polarization degree, polarization angle, and the Stokes parameters $Q/I$ and $U/I$, respectively. 
Solid lines with shaded areas represent the best fits of our decomposition procedure (see Section~\ref{sec_fit_polar} for a detail).}
\label{fig_polar}
\end{figure}

\subsection{Color Variations}
\label{sec_color}
Figure~\ref{fig_color} shows the color variations ($B-V$, $B-R$, and $B-J$) of 3C 273 with the $V$-band magnitude using the SMARTS data (\citealt{Bonning2012}).
All the three color indices vary with a complicated, time-dependent behaviour. 
For the sake of comparison, we divide the light curves into three segments (JD$<$5500, 5500$<$JD$<$7000,
and JD$>$7000) and plot the corresponding color indices in the right three panels of Figure~\ref{fig_color}. For the period of JD$<$5500, the color indices 
increase as the $V$-band magnitude decreases, indicating that 3C~273 become redder when brighter. By contrast, for the periods of  5500$<$JD$<$7000
and JD$>$7000, the variation behaviours are conversed, namely, 3C~273 becomes bluer when brighter. The differences between the periods of  5500$<$JD$<$7000
and JD$>$7000 are 1) the typical values of the color indices are not the same, as can be seen in the leftmost panel of Figure~\ref{fig_color}; and 2) 
the slopes of the color indices with the $V$-band magnitude are also not the same.

A number of previous studies had also investigated the color variability of 3C~273 on various time scales (e.g., \citealt{Dai2009, Ikejiri2011, Fan2014, Xiong2017, Zeng2018}).
Those studies basically found that the color variability of 3C~273 seems to transit between the bluer-when-brighter and redder-when-brighter trends,
plausibly depending on the observed epochs and brightness states. Our results are generally consistent with those reported behaviours.

There is consensus from large AGN samples that radio-quiet AGNs generally exhibit the bluer-when-brighter trend (e.g., 
\citealt{Schmidt2012,Ruan2014, Guo2016}), whereas in radio AGNs, both the bluer-when-brighter and redder-when-brighter trends 
are observed (e.g., \citealt{Gu2006, Rani2010, Bian2012}). The above observations imply that sole disk variability
cannot explain the complicated color variation behaviours in 3C 273.

\subsection{Polarization Variations}
\label{sec_polar}
Figure~\ref{fig_polar} plots the optical polarization degree and polarization angle of 3C 273 (5000-7000\AA) measured by the 
Steward Observatory spectropolarimetric monitoring project (\citealt{Smith2009}). 
Similar with normal radio-quiet AGNs (\citealt{Stockman1984, Brindle1990, Marin2014, Hutsemekers2018}), 3C 273 overall exhibits 
a low-level optical polarization of $p\sim$0.2\% 
in average, with the maximum up to $p\gtrsim0.6\%$. However, the polarization strikingly
undergoes large variations with $\Delta p/p>1$, a characteristic typically observed in blazar-like AGNs (\citealt{Impey1989}).
Meanwhile, the polarization angle also varies mildly with time. 

For normal radio-quiet AGNs, low-level polarizations are generally believed to originate from scattering of dust grains in torus or free electrons 
distributed somewhere in AGNs (\citealt{Stockman1984, Smith2002, Goosmann2007}, and references therein). The such resulting (linear) 
polarizations are not expected to show 
large variability in amount or orientation (\citealt{Rudy1983, Stockman1984}), which is generally supported by polarization observations for normal 
radio-quiet AGNs (e.g, \citealt{Stockman1984, Berriman1990, Brindle1990, Marin2014}). In this sense, the variable polarization 
degree and angle shown in Figure~\ref{fig_polar} directly
indicates non-negligible contributions of synchrotron emissions from the jet at optical wavelength (\citealt{Impey1989}).

\subsection{Detrending the Optical Continuum Emission}
\label{sec_detrend}
\begin{figure}
\centering
\includegraphics[width=0.46\textwidth]{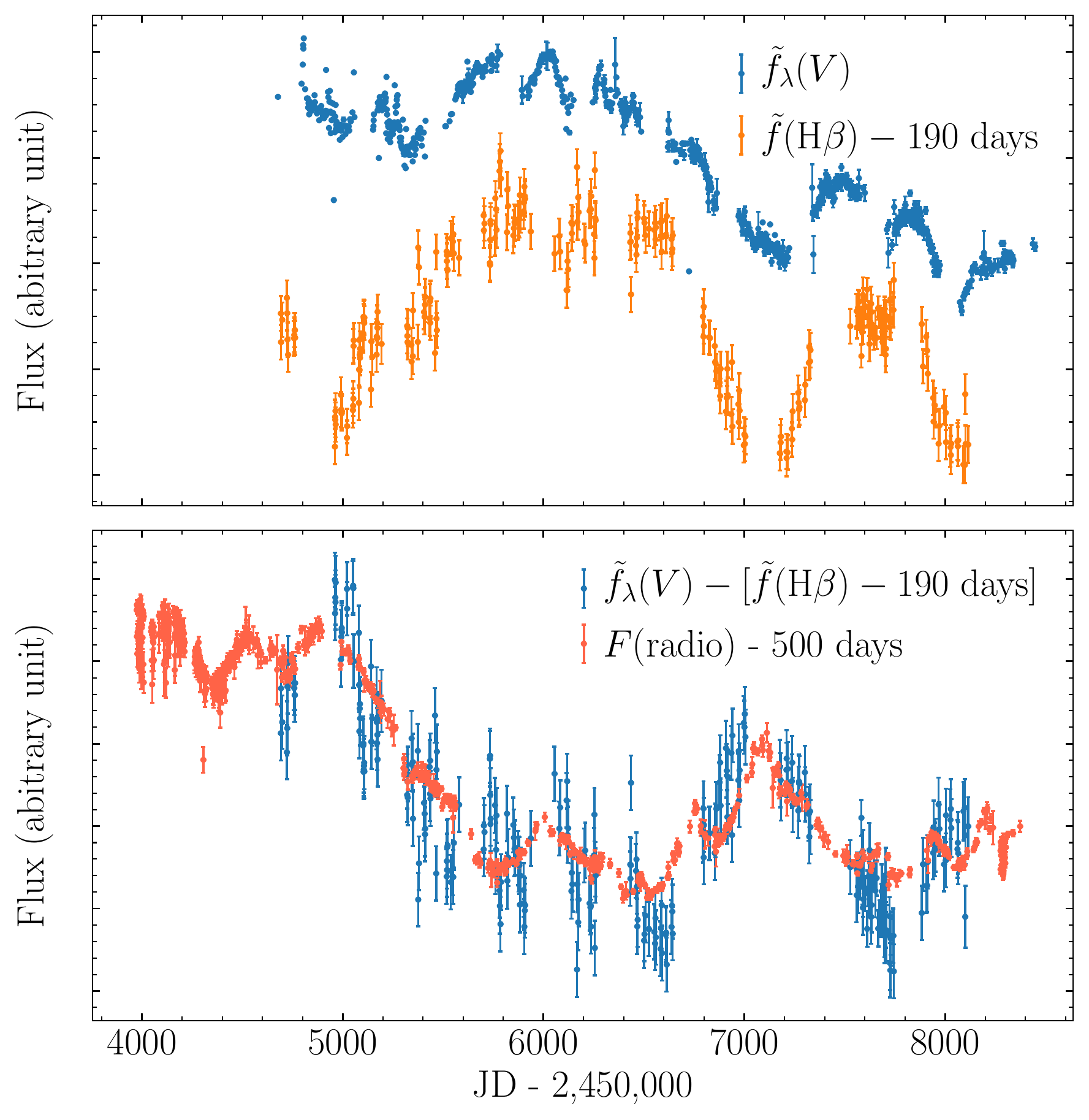}
\caption{(Top) The scaled $V$-band light curve denoted by $\tilde{f}(V)$ and scaled and shifted H$\beta$ light curve denoted by $\tilde{f}({\rm H\beta})$. 
(Bottom) The residuals $\tilde{f}(V) - \tilde{f}({\rm H\beta})$, on which superposed is the scaled radio 15 GHz light curve denoted by $F(\rm radio)$.
Note that $\tilde{f}({\rm H\beta})$ and $F(\rm radio)$ are additionally shifted backward by 190 and 500 days, respectively.
See \ref{sec_detrend} for the definitions of $\tilde{f}(V)$ and $\tilde{f}({\rm H\beta})$.}
\label{fig_detrend}
\end{figure}

As illustrated in Figure~\ref{fig_comp}, the optical light curve displays an extra long-term trend compared to the UV and H$\beta$ light curves.
Also, through simple shifting and scaling, the variation patterns of the UV and H$\beta$ light curves are well matched. In consideration of the 
high cadence of the H$\beta$ light curve, below we use it as a proxy for the UV light curve. We artificially scale and shift the  H$\beta$ light curve as 
\begin{equation}
\tilde{f}({\rm H}\beta) = 2.5 \times \frac{F({\rm H}\beta)}{\bar{F}({\rm H}\beta)} - 1.8,
\end{equation}
where $\bar{F}({\rm H}\beta)$ is the mean of the H$\beta$ light curve. The numbers in the above equation is chosen
for the purpose of illustration and do not have special meanings.
We then subtract $\tilde{f}({\rm H}\beta)$ from the scaled $V$-band light curve 
\begin{equation}
\tilde{f}(V) = \frac{F(V)}{\bar{F}(V)}, 
\end{equation}
and obtain the residuals
\begin{equation}
\Delta \tilde{f} =  \tilde{f}(V) - \tilde{f}({\rm H}\beta),
\end{equation}
where $\bar{F}(V)$ is the mean of the $V$-band light curve. Figure~\ref{fig_detrend} plots $\tilde{f}({\rm H}\beta)$ and
$\tilde{f}(V)$ in the top panel and $\Delta \tilde{f}$ in the bottom panel. It is remarkable that the variation pattern of the 
residual light curve highly resembles that of the radio light curve. To guide the eye, we also superpose the scaled radio light curve 
in the bottom panel of Figure~\ref{fig_detrend}.  After shifting backward about 500 days, the radio light curve well matches the 
residual light curve. This strongly suggests that the jet contamination is a plausible origin of the extra long-term trend
in the optical light curve.

In the above, we scale and shift the light curves artificially for illustration purpose. In reality, the relations among these 
light curves are by no means such simple, for example, according to line reverberation mapping scenario,
the H$\beta$ emission is linked to the UV continuum by convolving with a transfer function (e.g., \citealt{Peterson1993}).
Below, we develop a framework to untangle the optical jet and disk emissions in a rigorous mathematical foundation.

\begin{deluxetable*}{cccl}
\tablecolumns{4}
\tabletypesize{\footnotesize}
\tablecaption{Free Parameters. \label{tab_param}}
\tablehead{
\colhead{Parameter}  &
\colhead{Prior}      &
\colhead{Range}      &
\colhead{Definition} 
}
\startdata
$\sigma_{\rm d}$      & Logarithmic  &  ($10^{-5}$, 1.0)      &  Long-term standard deviation of the optical disk emission\\
$T_{\rm d}$           & Logarithmic  &  (1.0, $10^5$) days    &  Characteristic damping time scale of the optical disk emission\\
$\sigma_{\rm j}$      & Logarithmic  &  ($10^{-5}$, 1.0)      &  Long-term standard deviation of the optical jet emission\\
$T_{\rm j}$           & Logarithmic  &  (1.0, $10^5$) days    &  Characteristic damping time scale of the optical jet emission\\
$f_{\rm H\beta}$      & Logarithmic  &  ($10^{-3}$, $10^{3}$) &  Amplitude of the Gaussian transfer function for the H$\beta$ light curve\\
$\tau_{\rm H\beta}$   & Uniform      &  (0, 300) days         &  Mean time delay of the Gaussian transfer function for the H$\beta$ light curve \\
$\omega_{\rm H\beta}$ & Logarithmic  &  (0, 1000) days        &  Standard deviation of the Gaussian transfer function for the H$\beta$ light curve\\
$f_{\rm j}$           & Logarithmic  &  ($10^{-3}$, $10^{3}$) &  Amplitude of the Gaussian transfer function for the 15 GHz radio light curve\\
$\tau_{\rm j}$        & Uniform      &  (0, 1000) days        &  Mean time delay of the Gaussian transfer function for the 15 GHz radio light curve\\
$\omega_{\rm j}$      & Logarithmic  &  (0, 1000) days        &  Standard deviation of the Gaussian transfer function for the 15 GHz radio light curve\\
$f_{q}$               & Uniform      &  (0, 1)                &  Ratio of the mean of the disk emission to that of the total optical emission \\
$\bar{p}_{\rm j}$     & Uniform      &  (0, 2)\%              &  Averaged polarization degree of the optical jet emission\\
$\bar{\theta}_{\rm j}$& Uniform      &  (0, 180$^\circ$)      &  Averaged polarization angle of the optical jet emission
\enddata
\tablecomments{The prior ranges for $\sigma_{\rm d}$, $\sigma_{\rm j}$, $f_{\rm H\beta}$, and $f_{\rm j}$ are 
assigned in terms of the mean fluxes of all the light curves normalized to unity. A ``logarithmic'' prior means a uniform prior 
for the logarithm of the parameter. In real calculations, parameters with logarithmic priors are reparameterized by their logarithms.}
\end{deluxetable*}

\section{Untangling the Jet and Disk Emissions}
\label{sec_bayes}

\begin{figure*}
\centering 
\includegraphics[width=1.0\textwidth]{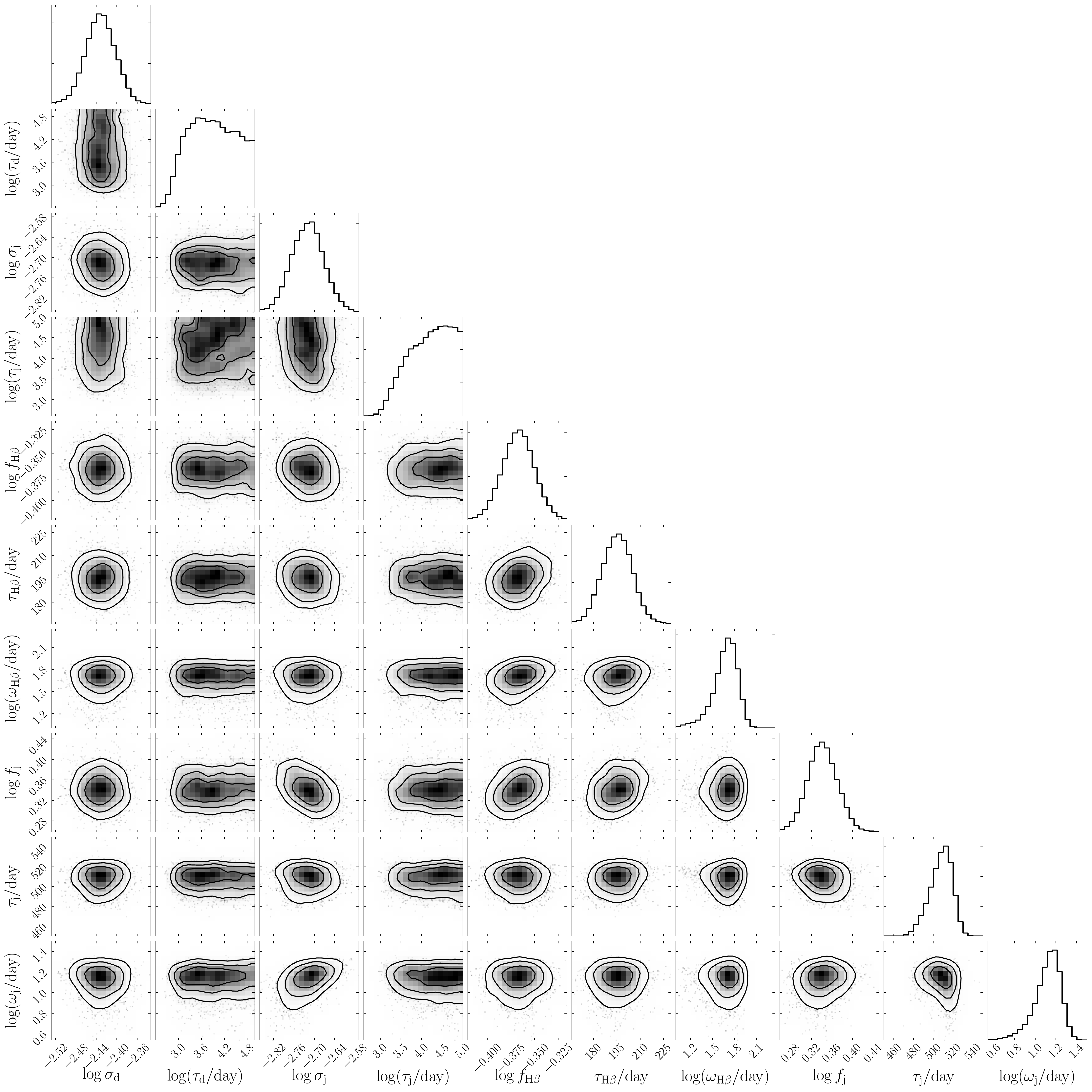}
\caption{One- and two-dimensional distributions of the free parameters. The contours are at 1$\sigma$, 1.5$\sigma$, and 2$\sigma$ levels.
This figure is made using the Python module \texttt{corner.py} (\citealt{Foreman2016}). }
\label{fig_hist}
\end{figure*}

\subsection{Basic Equations}
The observed optical emission is deemed to be a combination of the disk and jet emissions, namely,
\begin{equation}
C_{\rm t}(t) = C_{\rm d}(t) + C_{\rm j}(t),
\label{eqn_ct}
\end{equation}
where the subscripts ``t'', ``d'', and ``j'' represent the total, disk, and jet emissions, respectively.
According to the generic jet scenario (e.g., \citealt{Marscher1985, Turler2000}), perturbations propagate along jet from denser to less denser regions and 
the emitted photon energy gradually decreases from $\gamma$/X-ray, UV/optical to radio wavelengths. 
With this scenario, the jet emissions at optical and radio wavelengths are linked with a transfer function (also called delay map)  as
\begin{equation}
R_{\rm j}(t) = \int \Psi_{\rm j}(\tau)C_{\rm j}(t-\tau)d\tau,
\label{eqn_rj}
\end{equation}
where $R_{\rm j}(t)$ is flux at radio band and $\Psi_{\rm j}(\tau)$ is the transfer function
at the time delay $\tau$. 
The H$\beta$ emission line responds to the continuum emission from the accretion disk as (\citealt{Peterson1993})
\begin{eqnarray}
L_{\rm H\beta}(t) &=& \int \Psi_{\rm H\beta}(\tau)C_{\rm d}(t-\tau)d\tau,
\label{eqn_hb}
\end{eqnarray}
where $L_{\rm H\beta}(t)$ is the line flux and $\Psi_{\rm H\beta}(\tau)$ is the transfer function of the BLR.

By appropriately solving Equations (\ref{eqn_ct}-\ref{eqn_hb}), we can separate the optical emissions from the disk $C_{\rm d}(t)$ and the jet $C_{\rm j}(t)$.
However, this is challenging as the transfer functions $\Psi_{\rm j}$ and $\Psi_{\rm d}$ are fully unknown. Below we proceed with 
several simple, but physically reasonable assumptions and show how to recover the disk and jet emissions as well as the two transfer functions 
based on the framework of linear reconstruction of irregularly sampled time series outlined by \cite{Rybicki1992}.

First, we assume that time variations of the disk and jet emissions are described by two independent damped random walk (DRW) processes.
DRW processes have been applied to various time series data with their capability of capturing main variation features (e.g. \citealt{Kelly2009, 
Zu2011, Pancoast2014, Li2014, Li2016, Li2018}).
A DRW process is a stationary Gaussian process and its covariance between times $t_k$ and $t_m$ damps exponentially with the time difference $\Delta t=t_m-t_k$.
As such, the auto-covariance functions of $C_{\rm d}(t)$ and $C_{\rm j}(t)$ are given by
\begin{eqnarray}
S_{\rm dd}(\Delta t) &=& \left\langle C_{\rm d}(t_k) C_{\rm d}(t_m) \right\rangle = \sigma_{\rm d}^2 \exp\left(-\frac{|\Delta t|}{T_{\rm d}}\right),\label{eqn_sdd}\\
S_{\rm jj}(\Delta t) &=& \left\langle C_{\rm j}(t_k) C_{\rm j}(t_m) \right\rangle = \sigma_{\rm j}^2 \exp\left(-\frac{|\Delta t|}{T_{\rm j}}\right),\label{eqn_sjj}
\end{eqnarray}
where the angle brackets represent the statistical ensemble average, $T_{\rm d}$ and $T_{\rm j}$ are the characteristic damping time scales, 
and $\sigma_{\rm d}$ and $\sigma_{\rm j}$ are the long-term standard deviations of the disk and jet emissions, respectively.  

From the fundamental plane of black hole activity (e.g., \citealt{Merloni2003}), we know that there must be somehow disk-jet connection over the lifetime of 
the black hole activity, which is typically on the order of million years (e.g., \citealt{Martini2004}). 
Nevertheless, the assumption that $C_{\rm d}(t)$ and $C_{\rm j}(t)$ are independent still stands reasonable in the sense that we are only concerned with  variations 
on a much shorter timescale ($\sim$ years), which are most likely driven by independent fluctuations/perturbations in the disk and jet. 
Since $C_{\rm d}(t)$ and $C_{\rm j}(t)$ are independent, their covariance is simply zero.
The auto-covariance function of $C_{\rm t}(t)$ is thereby
\begin{equation}
S_{\rm tt}(\Delta t) = \left\langle C_{\rm t}(t_k) C_{\rm t}(t_m) \right\rangle = S_{\rm dd}(\Delta t) + S_{\rm jj}(\Delta t).
\end{equation}

Second, for simplicity, we assume that the transfer functions $\Psi_{\rm j}(\tau)$ and $\Psi_{\rm d}(\tau)$
are parameterized by Gaussians (see also Section \ref{sec_tf} below),
\begin{equation}
\Psi_{\rm H\beta}(\tau) = \frac{f_{\rm H\beta}}{\sqrt{2\pi}\omega_{\rm H\beta}} \exp\left[ - \frac{(\tau - \tau_{\rm H\beta})^2}{2\omega^2_{\rm H\beta}}\right],
\label{eqn_tfd}
\end{equation}
and 
\begin{equation}
\Psi_{\rm j}(\tau) = \frac{f_{\rm j}}{\sqrt{2\pi}\omega_{\rm j}} \exp\left[ - \frac{(\tau - \tau_{\rm j})^2}{2\omega^2_{\rm j}}\right],
\label{eqn_tfj}
\end{equation}
where ($f_{\rm H\beta}$, $\tau_{\rm H\beta}$, $\omega_{\rm H\beta}$) and ($f_{\rm j}$, $\tau_{\rm j}$, $\omega_{\rm j}$) are free parameters. 
By this definition, $\tau_{\rm H\beta}$ and $\tau_{\rm j}$ represent the time delays of the H$\beta$ emission and radio emission relative to the optical emission, respectively. 
With Gaussian transfer functions, the covariances among $C_{\rm t}(t)$, $R_{\rm j}(t)$, and $L_{\rm H\beta}(t)$ can be expressed analytically by the aid of error function (see 
Appendix B).

\begin{figure}
\centering 
\includegraphics[width=0.48\textwidth]{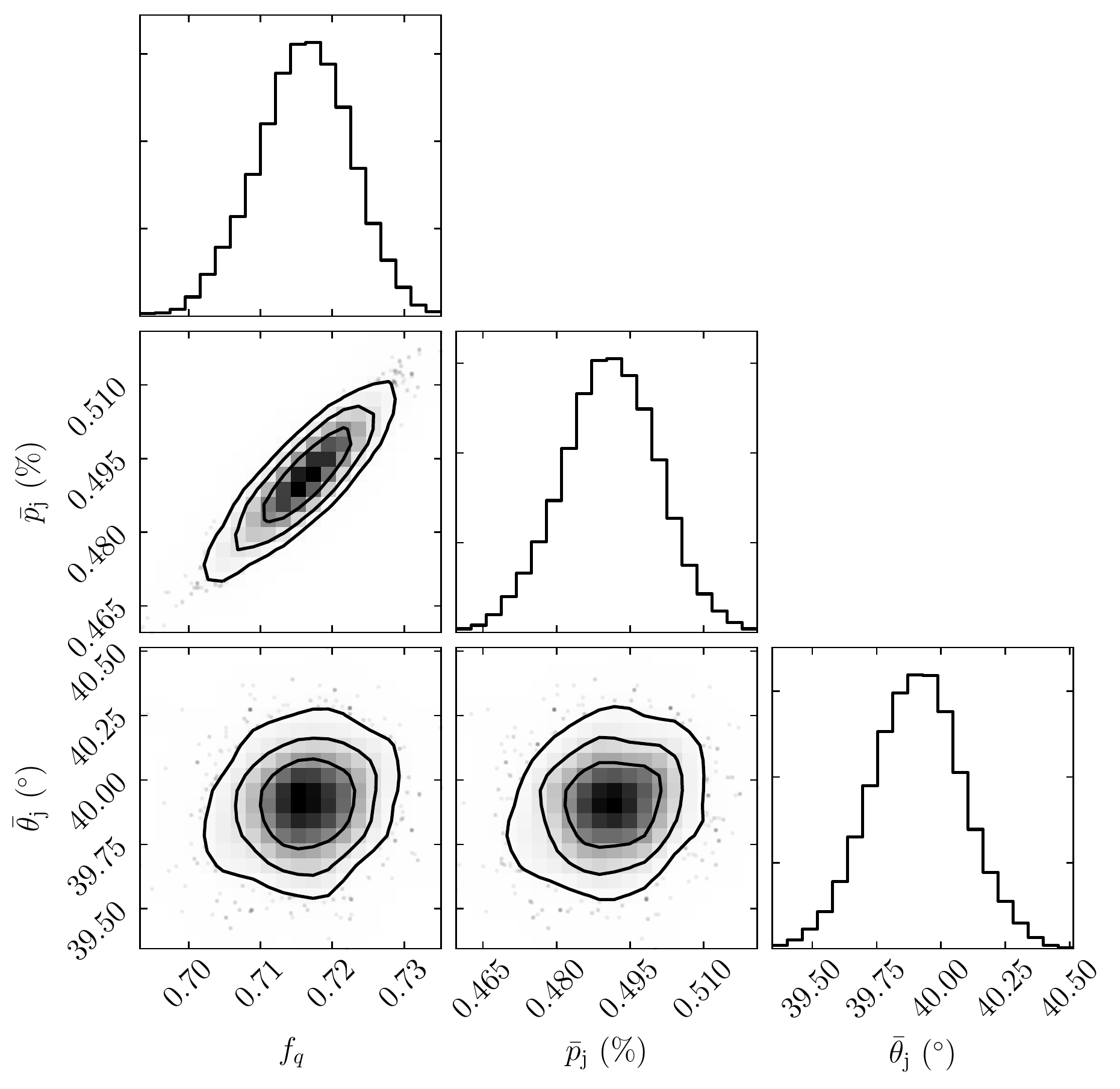}
\caption{The posterior distributions of $f_q$, $\bar p_{\rm j}$, and $\bar \theta_{\rm j}$. The contours are at 1$\sigma$, 1.5$\sigma$, and 2$\sigma$ levels.
This figure is made using the Python module \texttt{corner.py} (\citealt{Foreman2016}).}
\label{fig_mean_polar}
\end{figure}

\subsection{Bayesian Inference}
The observation data at hand are the optical continuum $C_{\rm t}(t)$, the radio flux $R_{\rm j}(t)$, the H$\beta$ emission line flux $L_{\rm H\beta}(t)$, and their 
respective associated measurement uncertainties. For brevity, we join [$C_{\rm t}(t)$, $R_{\rm j}(t)$, $L_{\rm H\beta}(t)$] to a vector and denote it by $\mathbi{y}$. 
By assuming that the measurement noises are Gaussian, the likelihood probability for $\mathbi{y}$ is (\citealt{Rybicki1992,Zu2011, Li2013})
\begin{eqnarray}
P(\mathbi{y}|\boldsymbol{\theta})& = &\frac{1}{\sqrt{(2\pi)^{m}|\mathbi{C}||\mathbi{E}^T\mathbi{C}^{-1}\mathbi{E}|}}\nonumber\\
& \times&\exp\left[-\frac{(\mathbi{y}-\mathbi{E\hat q})^T\mathbi{C}^{-1}(\mathbi{y}-\mathbi{E\hat q})}{2}\right],
\label{eqn_likeli}
\end{eqnarray}
where $\boldsymbol{\theta}$ denotes all the free parameters, $m$ is the total number of data points in $\mathbi{y}$,
$\mathbi{C}=\mathbi{S}+\mathbi{N}$, $\mathbi{S}$ is the covariance matrix of $\mathbi{y}$, $\mathbi{N}$ is the covariance matrix of the measurement noises, 
$\mathbi{\hat q}$ is a vector with three entries that represent the best estimate for the means of $\mathbi{y}$,
\begin{equation}
\mathbi{\hat q} = (\mathbi{E}^T\mathbi{C}^{-1}\mathbi{E})^{-1}\mathbi{E}^{T}\mathbi{C}^{-1} \mathbi{y},
\label{eqn_q}
\end{equation}
and $\mathbi{E}$ is a $3\times m$ matrix with entries of $(1, 0, 0)$ for the optical continuum data points, 
$(0, 1, 0)$ for the radio data points, and $(0, 0, 1)$ for the H$\beta$ flux data points.

The posterior probability for the free parameters $\boldsymbol{\theta}$ is 
\begin{equation}
P(\boldsymbol{\theta}|\mathbi{y}) = \frac{P(\mathbi{y}|\boldsymbol{\theta})P(\boldsymbol{\theta})}{P(\mathbi{y})},
\label{eqn_post}
\end{equation}
where $P(\mathbi{y})$ is the marginal likelihood and $P(\boldsymbol{\theta})$ is the prior probability of the free parameters.
The prior probabilities for $\tau_{\rm H\beta}$ and $\tau_{\rm j}$ are set to be uniform and for the other parameters are set to be 
a logarithmic prior. Table~\ref{tab_param} lists the priors for all the free parameters. 
We employ the diffusive nested sampling technique\footnote{We wrote a C version code \texttt{CDNest} for the 
diffusive nested sampling algorithm proposed by \cite{Brewer2011}. The code is adapted with the standardized message passing interface to 
implement on parallel computers/clusters. The code is publicly available at \url{https://github.com/LiyrAstroph/CDNest}.} (\citealt{Brewer2011}) to explore the posterior probability
and construct posterior samples to determine the best estimates and 
uncertainties for the free parameters. With the best estimated parameters, a reconstruction of $\mathbi{y}$ is given by 
\begin{equation}
\mathbi{\hat y} = \mathbi{SC}^{-1}(\mathbi{y-E\hat q}) + \mathbi{E\hat q}.
\label{eqn_rec}
\end{equation}

\begin{deluxetable}{lc}
\renewcommand\arraystretch{1.2}
\tablecolumns{2}
\tabletypesize{\footnotesize}
\tablecaption{Inferred Parameter Values and Uncertainties.\label{tab_value}}
\tablehead{
\colhead{Parameter\qquad\qquad\qquad\qquad\qquad\qquad}   &
\colhead{\quad\qquad\qquad Value\qquad\qquad\qquad\quad}  
}
\startdata
$\log\sigma_{\rm d}$                   & $-2.430_{-0.028}^{+0.028}$\\
$\log(T_{\rm d}/{\rm day})$            & $3.92_{-0.67}^{+0.71}$\\
$\log\sigma_{\rm j}$                   & $-2.717_{-0.046}^{+0.042}$\\
$\log(T_{\rm j}/{\rm day})$            & $4.28_{-0.61}^{+0.49}$\\
$\log f_{\rm H\beta}$                  & $-0.367_{-0.016}^{+0.016}$ \\
$\tau_{\rm H\beta}/{\rm day}$          & $195.4_{-9.2}^{+8.6}$\\
$\log(\omega_{\rm H\beta}/{\rm day})$  & $1.69_{-0.20}^{+0.12}$\\
$\log f_{\rm j}$                       & $0.341_{-0.026}^{+0.028}$\\
$\tau_{\rm j}/{\rm day}$               & $509.2_{-12.8}^{+8.0}$ \\
$\log(\omega_{\rm j}/{\rm day})$       & $1.143_{-0.130}^{+0.084}$\\
$f_q$                                  & $0.716_{-0.006}^{+0.006}$\\
$\bar{p}_{\rm j}/\%$                   & $0.491_{-0.009}^{+0.009}$\\
$\bar{\theta}_{\rm j}/^\circ$          & $39.91_{-0.17}^{+0.17}$
\enddata
\tablecomments{Parameter values are determined from the medians of the posterior probability distributions
and the uncertainties represent the 68.3\% confidence intervals. The values of $f_q$, $\bar{p}_{\rm j}$, and 
$\bar{\theta}_{\rm j}$ are given by fixing the interstellar polarization degree and angle to $\theta_\star=0.15\%$
and $\theta_\star=70^\circ$.}
\end{deluxetable}

It is worth stressing that the likelihood probability $P(\mathbi{y}|\boldsymbol{\theta})$
in Equation (\ref{eqn_likeli}) are fully determined without involving extra parameters 
that we will introduce below to untangle the mean fluxes of the disk and jet emissions. 
Therefore, at this point we can already obtain the posterior samples for the free parameters defined above
by optimizing the posterior probability $P(\boldsymbol{\theta}|\mathbi{y})$
in Equation (\ref{eqn_post}). This means that we can ``detrend'' the light curves to
determine the time delays ($\tau_{\rm H\beta}$ and $\tau_{\rm j}$) without invoking the need of 
knowing the mean fluxes of individual emission components.

\begin{figure}
\centering 
\includegraphics[width=0.48\textwidth]{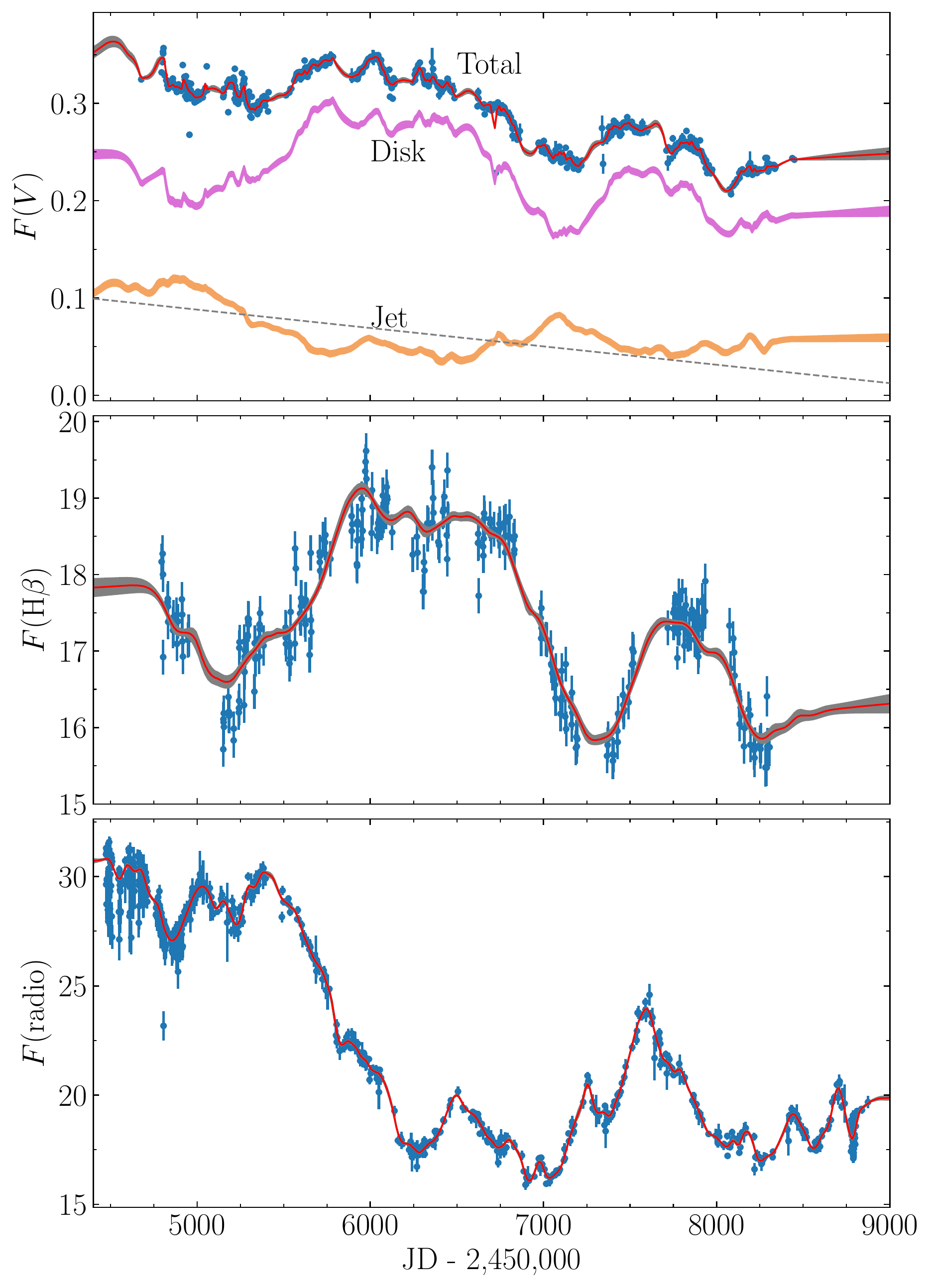}
\caption{Best fits to the $V$-band, H$\beta$, and radio light curves. Points with error bars represent the observed data
and red lines with shaded areas represent the best fits. In the top panel, superposed are the decomposed disk and jet light curves. 
Grey dashed line represents the linear polynomial used to detrend the continuum light curve by \cite{Zhang2019}, 
which is shifted vertically to guide the eye. }
\label{fig_recon}
\end{figure}

\subsection{Including the Polarization Data}
\label{sec_fit_polar}
In Equation~(\ref{eqn_rec}), there involve the means $\mathbi{\hat q}$ of the light curves, which are calculated by Equation (\ref{eqn_q}). 
This implies that when reconstructing $C_{\rm j}(t)$ and 
$C_{\rm d}(t)$, their means are degenerated since we only know the sum of $C_{\rm j}(t)$ and $C_{\rm d}(t)$, namely $C_{\rm t}(t)$. 
We define a free parameter $f_{q}$ to denote the ratio of the mean of the disk emission to
that of the total optical emission. With this definition, we reconstruct $C_{\rm j}(t)$ and $C_{\rm d}(t)$ as 
\begin{equation}
\label{eqn_yd}
\mathbi{\hat y}_{\rm d} = \mathbi{S_{\rm d*}C}^{-1}(\mathbi{y-E\hat q}) + f_q\mathbi{E}\hat q_{\rm t},
\end{equation}
and
\begin{equation}
\label{eqn_yj}
\mathbi{\hat y}_{\rm j} = \mathbi{S_{\rm j*}C}^{-1}(\mathbi{y-E\hat q}) + (1-f_q)\mathbi{E}\hat q_{\rm t},
\end{equation}
where $\hat q_{\rm t}$ is the mean of $C_{\rm t}(t)$, $\mathbi{S_{\rm d*}}$
and $\mathbi{S_{\rm j*}}$ are covriances of $C_{\rm d}(t)$ and $C_{\rm j}(t)$
with the observed light curves $\mathbi{y}$ at reconstructed times, respectively.

The restriction that both  $C_{\rm j}(t)$ and $C_{\rm d}(t)$ must be positive can constrain $f_q$ to a generic range of $f_q\sim (0.18-0.83)$.
To further constrain the value of $f_q$, we resort to the polarization data shown in Figure~\ref{fig_polar}. 
The observed optical polarization is deemed to be a combination of interstellar 
polarization and polarizations of the disk and jet emissions. 
As usual, we depict polarization using the Stokes parameters ($I, Q, U, 0$) ($V=0$ for linear polarization; e.g., \citealt{Chandrasekhar1960}).
The optical polarization of 3C 273 is then given by 
\begin{eqnarray}
I &=& C_{\rm t} = C_{\rm d} + C_{\rm j},\label{eqn_polar1}\\
Q &=& C_{\rm d} p_{\rm d} \cos(2\theta_{\rm d}) + C_{\rm j} p_{\rm j} \cos(2\theta_{\rm j}) + C_{\rm t} p_{\star} \cos(2\theta_{\star}),\label{eqn_polar2}\\
U &=& C_{\rm d} p_{\rm d} \sin(2\theta_{\rm d}) + C_{\rm j} p_{\rm j} \sin(2\theta_{\rm j})\label{eqn+polar3}
+ C_{\rm t} p_{\star} \sin(2\theta_{\star})\label{eqn_polar3},
\end{eqnarray}
where $p$ and $\theta$ represent polarization degree and position angle, and 
the subscripts ``d'', ``j'', and ``$\star$'' represent disk, jet, and interstellar polarization, respectively. 
The polarization degree and angle of the total emission are then calculated as 
\begin{eqnarray}
p &=& \frac{\sqrt{U^2+Q^2}}{I},\\
\theta &=& \frac{1}{2}\tan^{-1}\left(\frac{U}{Q}\right).
\end{eqnarray}
It is expected that the 
interstellar polarization is constant while the polarizations of the disk and jet 
emissions might vary with time.  In addition, there might be a time delay between the polarized and unpolarized disk emissions
depending on the location of scattering materials (e.g. \citealt{Gaskell2012, Rojas2020}).

The interstellar polarization degree can be estimated by an approximation 
$p_\star(\%) = 0.3 E(B-V)$ (\citealt{Rudy1983}). The reddening for 3C 273 is $E(B-V)=0.05$ mag (\citealt{Wu1977}), leading to an interstellar 
polarization of $p_\star=0.15\%$. This crude estimate is generally consistent with the observational constraints of 
$\sim0.17\%$ by \cite{Impey1989} and $\sim0.13\%$ by \cite{Smith1993}. The correspondingly estimated position angle of the interstellar polarization 
is $\theta_\star\sim80^\circ$ by \cite{Impey1989} and $\theta_\star\sim60^\circ$ by \cite{Smith1993}.
We hereafter use the medians $p_\star=0.15\%$ and $\theta_\star=70^\circ$ as the fiducial values for the interstellar polarization.

For disk emissions, there have been several studies that calculated polarizations arising from scattering by electrons in disk atmospheres 
(\citealt{Chandrasekhar1960, Phillips1986, Agol1996, Beloborodov1998, Li2009}) or by dust grains/electrons distributed beyond the disk 
(such as in torus or polar regions; \citealt{Wolf1999, Goosmann2007, Rojas2018}). All those calculations showed that the polarization
of disk emissions strongly depends on the view inclination of the disk. As the inclination approaches face-on ($i\rightarrow0$), 
the polarization decreases to zero 
because of symmetry. We note that observations of the superluminal motion of the radio jet of 3C 273 yielded an inclination angle 
of $i\sim 10^\circ$ (\citealt{Abraham1999,Savolainen2006, Jostad2017}). If we assume that the jet is aligned with the disk's rotation axis, 
the inclination angle 
of the disk would be also $\sim 10^\circ$. For an optically thick disk with pure scattering, such an inclination results in a quite low 
degree of polarization $\sim$0.05\% \citep[Section X]{Chandrasekhar1960}. Once taking into account photon absorption, this degree of 
polarization will be further reduced as absorption tends to 
destroy the polarization of photons. The calculations of \cite{Beloborodov1998} showed that the presence of a fast wind flowing away from
the accretion disk will alter the disk's intrinsic polarization and produce a high polarization degree at large inclination angles.
However, at low inclination angles, the effect of a disk wind is again insignificant (see Figure 2 therein). 
For cases of polarization caused by scattering of dust grains/electrons distributed beyond the accretion disk,
the studies of \cite{Wolf1999} and \cite{Goosmann2007} also generally indicated a pretty low degree of polarization at inclination 
of $i\sim10^\circ$. Taken the above together, we neglect the polarization of disk emissions by default ($p_{\rm d}=0$)
and below we will discuss the influences on our conclusions if the polarization of disk emissions is not negligible.

Equations~(\ref{eqn_polar1}-\ref{eqn_polar3}) implies that the time-dependent ratio $C_{\rm j}(t)/C_{\rm t}(t)$ does contribute to the observed 
polarization variability. In generic, both the polarization degree $p_{\rm j}$ and angle $\theta_{\rm j}$
can change with time. In such a case, Equations~(\ref{eqn_polar1}-\ref{eqn_polar3}) cannot uniquely determine $f_{\rm q}$ and time-dependent 
$p_{\rm j}$ and $\theta_{\rm j}$. As illustrated below, the global variation patterns of the observed polarization degree
(plotted in Figure~\ref{fig_polar}) are generally similar to these of the ratio $C_{\rm j}(t)/C_{\rm t}(t)$, except for patterns in 
short timescales within about one year. This motivates us to make a assumption that the global variations of the observed 
polarizations are mainly contributed from the ratio $C_{\rm j}(t)/C_{\rm t}(t)$ whereas the intrinsic variations
of the jet polarizations are mainly responsible for the short-timescale variations. As such, by using Equations~(\ref{eqn_polar1}-\ref{eqn_polar3}) 
to fit the observed $Q$ and $U$ (see Figure~\ref{fig_polar}), we can uniquely determines $f_{\rm q}$ and the averaged values of $p_{\rm j}(t)$ 
and $\theta_{\rm j}(t)$, which we denote as $\bar p_{\rm j}$ and $\bar \theta_{\rm j}$, respectively.  The priors for $f_q$, $\bar p_{\rm j}$, 
and $\bar\theta_{\rm j}$ are listed in Table~\ref{tab_param}.

We stress again that the likelihood probability $P(\mathbi{y}|\boldsymbol{\theta})$ in Equation~(\ref{eqn_likeli}) 
does not involve $f_q$, $\bar p_{\rm j}$, and $\bar \theta_{\rm j}$. Their values can thereby be determined after obtaining the 
posterior parameter samples from the posterior 
probability $P(\boldsymbol{\theta}|\mathbi{y})$ in Equation~(\ref{eqn_post}).
Moreover, we can find from Equations~(\ref{eqn_yd}-\ref{eqn_yj}) that the parameter $f_q$ only
controls the means of $C_{\rm d}(t)$ and $C_{\rm j}(t)$, but does not affect their variation patterns at all.
In other words, the variation pattern of the ratio $C_{\rm j}(t)/C_{\rm t}(t)$ (if regardless of its mean) is fully
determined from likelihood probability $P(\mathbi{y}|\boldsymbol{\theta})$, independent of $f_q$. 

\begin{figure}
\centering 
\includegraphics[width=0.48\textwidth]{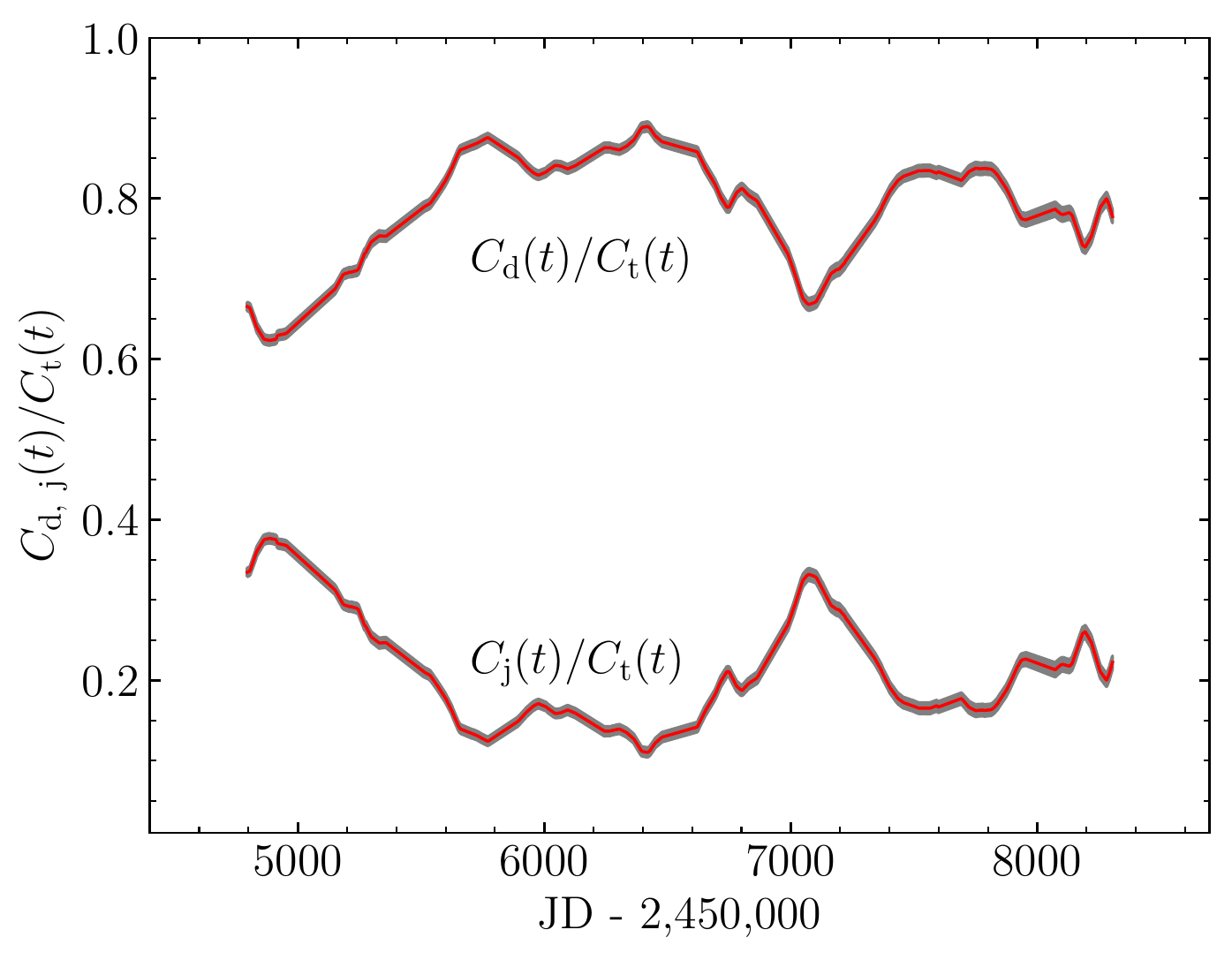}
\caption{The ratio of the decomposed optical disk and jet flux to the total optical flux. 
Shaded areas represent the uncertainties.}
\label{fig_ratio}
\end{figure}

\subsection{Results}
\subsubsection{Overview}
Figures~\ref{fig_hist} and \ref{fig_mean_polar} show the one- and two-dimensional distributions of the free parameters. 
The contours are plotted at $1\sigma$, $1.5\sigma$, and $2\sigma$ levels. Table~\ref{tab_value} summarizes the best estimated parameter values
determined from the medians of the posterior probability distributions and uncertainties determined from the 68.3\% confidence intervals.
The time delay of the H$\beta$ light curve 
with respect to the $V$-band light curve is $\tau_{\rm H\beta}=194.9_{-9.8}^{+9.7}$ days (in observed frame), slightly larger than 
$\tau_{\rm H\beta}=170.0_{-14.0}^{+9.6}$ days reported by \cite{Zhang2019}, which used a linear fit to detrend the optical light curve.
The time delay of the radio 15 GHz light curve with respect to the $V$-band light curve is $\tau_{\rm j}=501.0_{-10.2}^{+7.5}$ days.
If regardless of the distance of the optical jet emission to the central black hole, the bulk of the 15 GHz emission is located 
at $c\tau_{\rm j}/(1+z)=0.36$~pc away from the black hole, where $c$ is the speed of light.
 
From Figure~\ref{fig_mean_polar}, we find that the averaged polarization degree of the jet emission at optical wavelengths
is $\bar{p}_{\rm j}\approx0.5\%$. This is consistent with the observed radio polarization $p\leqslant1\%$ in the radio core of 
3C 273 using the Very Long Baseline Array (\citealt{Attridge2001,Attridge2005}; see also \citealt{Hovatta2019}). Such a low-level 
polarization in the core of the jet was generally ascribed to the differential Faraday depolarization. 
The averaged polarization angle is $\bar\theta_{\rm j}\approx40^\circ$, which well agrees with 
the position angle ($\sim42^\circ$) of the jet structure of 3C 273 (e.g., \citealt{Roeser1991}).
Such an alignment between polarization angle and jet structure axis was commonly observed  
in blazars (\citealt{Rusk1985,Impey1989, Blinov2020} and references therein). 

In Figure~\ref{fig_recon}, we plot best fits to the $V$-band, H$\beta$, and radio light curves. The decomposed disk and 
jet light curves at $V$-band are also superposed in the top panel of Figure~\ref{fig_recon}. The $V$-band and radio 
light curves are well fitted. For the H$\beta$ light curve, although there are several detailed minor features (e.g., around JD 5250 and 8000) 
that cannot be reproduced, 
the main reconstructed variation patterns are in good agreement with observations. The discrepancies 
for these minor features may be ascribed to twofold reasons: 1) the assumed Gaussian transfer functions are simple 
so that the fitting is not expected to capture all the detailed features; 2) the 15 GHz radio emission of 3C 273 is core dominated, but 
there might be still a minor contamination from the large-scale jet (e.g., \citealt{Perley2017}), which is not correlated with the optical jet emission.
Nevertheless, as a zero-order approximation, our simple approach seems reasonable and enlightening.

For the sake of comparison, in the top panel of Figure~\ref{fig_recon}, we also plot the linear polynomial used to detrend 
the continuum light curve by \cite{Zhang2019}. Here, the polynomial is shifted vertically to align with the jet light curve. 
As can be seen, the slope of the polynomial is generally consistent with the declining trend of the jet light curve,
indicating the validity of our decomposition procedure.

Figure~\ref{fig_ratio} plots the obtained ratio of the optical disk and jet emission 
to the total emissions as a function of time, namely, $C_{\rm j}(t)/C_{\rm t}(t)$.
The jet contributes $\sim10\%$ of the optical emission at the minimum and $\sim40\%$ at the maximum. 
The best fits to the optical polarization degree, polarization angle,
and the Stokes parameters $Q/I$ and $U/I$ are plotted in Figure~\ref{fig_polar}.
As illustrated in Equations~(\ref{eqn_polar1}-\ref{eqn_polar3}), besides the interstellar polarization and the accretion disk's polarization, 
there are two contributions to the observed polarization variability: one is from 
$C_{\rm j}(t)/C_{\rm t}(t)$ and the other is from the jet itself $p_{\rm j}(t)$ and $\theta_{j}(t)$. 
By a visual inspection to Figures~\ref{fig_polar} and \ref{fig_ratio}, we can find 
the global variation structures in  $C_{\rm j}(t)/C_{\rm t}(t)$ generally match these of
the observed optical polarization degree (except in short timesscales).
Considering the fact that the variation patterns of $C_{\rm j}(t)/C_{\rm t}(t)$ (if regardless of its mean) 
does not depend on $f_q$, $\bar{p}_{\rm j}$, and $\bar{\theta}_{\rm j}$, the consistence in the global variation structures
reinforces our assumption that the global polarizations are mainly contributed from the ratio $C_{\rm j}(t)/C_{\rm t}(t)$.
Therefore, it is approximately viable to use averaged polarization degree $\bar p_{\rm j}$ and angle $\bar\theta_{\rm j}$
to decompose the optical light curves.
However, it is because of this assumption that we cannot reproduce all the detailed, rapid variability in polarizations
within timescales of months. Adding time-dependent perturbations to $\bar p_{\rm j}$ and $\bar \theta_{\rm j}$
would better fit the rapid polarization variability, but does not change the results of our calculations.

\begin{figure}
\centering 
\includegraphics[width=0.48\textwidth]{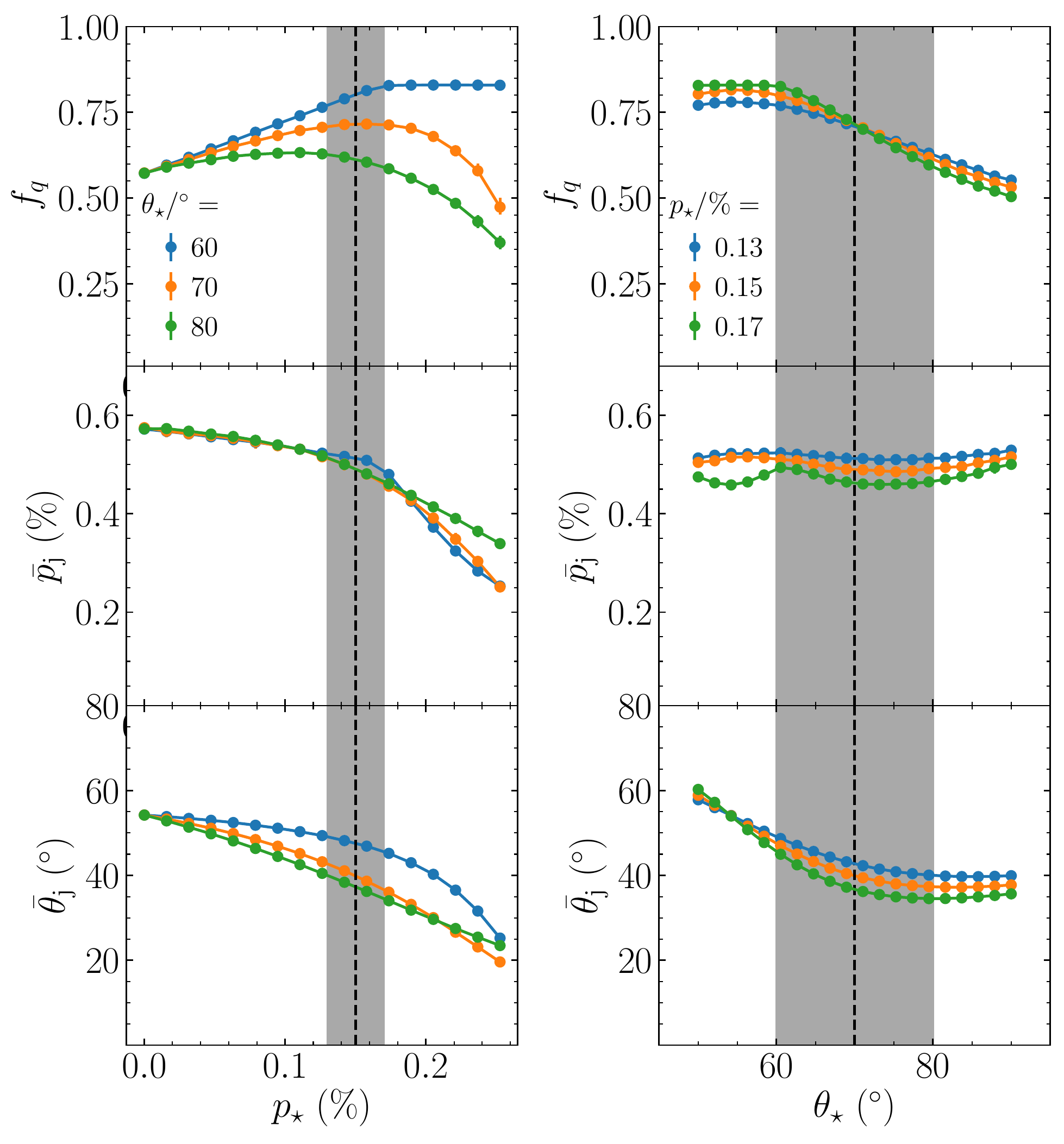}
\caption{The best estimated values of $f_{q}$, $\bar p_{\rm j}$, and $\bar \theta_{\rm j}$
for different input interstellar (left) polarization degree $p_\star$ and (right) polarization angle $\theta_\star$. 
Dashed vertical lines represent the fiducial values $p_\star=0.15\%$ and $\theta_\star=70^\circ$.
Shaded vertical bands represent the ranges between two previous observationally constrained interstellar polarization 
degree and angle ($p_\star=0.17\%$, $\theta_\star=80^\circ$) by \cite{Impey1989}
and ($p_\star=0.13\%$, $\theta_\star=60^\circ$) by \cite{Smith1993}.}
\label{fig_pstar}
\end{figure}

\begin{deluxetable}{ccccc}
\renewcommand\arraystretch{1.2}
\tablecolumns{5}
\tabletypesize{\footnotesize}
\tablecaption{Inferred Values of $f_q$, $\bar{p}_{\rm j}$, and  $\bar{\theta}_{\rm j}$ for 
different input interstellar polarization ($p_\star,\theta_\star$).\label{tab_polar}}
\tablehead{
\colhead{~~~$p_\star$~~~}   &
\colhead{~~~$\theta_\star$~~~} &
\colhead{~~~~~~~~~~~~$f_q$~~~~~~~~~~~~} &
\colhead{~~~~~~~~~~~~$\bar{p}_{\rm j}$~~~~~~~~~~~~} &
\colhead{~~~~~~~~~~~~$\bar{\theta}_{\rm j}$~~~~~~~~~~~~}\\
\colhead{(\%)}  &
\colhead{($^\circ$)} &
\colhead{}  &
\colhead{(\%)}  &
\colhead{($^\circ$)} 
}
\startdata
0.13 & 60 & $0.771_{-0.005}^{+0.005}$ & $0.523_{-0.010}^{+0.010}$ & $49.12_{-0.18}^{+0.19}$\\
0.13 & 80 & $0.626_{-0.009}^{+0.007}$ & $0.513_{-0.010}^{+0.009}$ & $40.00_{-0.13}^{+0.13}$\\
0.17 & 60 & $0.827_{-0.003}^{+0.002}$ & $0.491_{-0.007}^{+0.006}$ & $45.64_{-0.22}^{+0.24}$\\
0.17 & 80 & $0.591_{-0.010}^{+0.009}$ & $0.466_{-0.010}^{+0.010}$ & $34.56_{-0.13}^{+0.13}$
\enddata
\end{deluxetable}

\subsubsection{The Influences of the Interstellar Polarization}

In the above calculations, we fix the interstellar polarization degree $p_\star=0.15\%$ and polarization 
angle $\theta_\star=70^\circ$. In Figure~\ref{fig_pstar}, we show the dependence of the obtained 
$f_{q}$, $\bar p_{\rm j}$, and $\bar \theta_{\rm j}$ on the input interstellar polarization degree and 
angle. As described above, the previous observational constrains on the interstellar polarization along the direction
of 3C 273 include ($p_\star=0.17\%$, $\theta_\star=80^\circ$) by \cite{Impey1989}
and ($p_\star=0.13\%$, $\theta_\star=60^\circ$) by \cite{Smith1993}. 
In Table~\ref{tab_polar}, we list the resulting values of $f_{q}$, $\bar p_{\rm j}$, and $\bar \theta_{\rm j}$
for different pairs of the interstellar polarization ($p_\star,\theta_\star$).
By changing $p_\star$ from 0.13\% to 0.17\% with fixed $\theta_\star$, the influences on the obtained parameter values are 
minor. For $\theta_\star$ at a range of $(60^\circ-80^\circ)$, the resulting
$\bar p_{\rm j}$ almost has no change, while $f_q$ varies between $\sim$0.6 and $\sim$0.8 and $\bar \theta_{\rm j}$
varies between $\sim$35$^\circ$ and  $\sim$49$^\circ$.

\subsubsection{The Influences of Polarization of the Disk Emission}
\begin{figure}
\centering 
\includegraphics[width=0.28\textwidth]{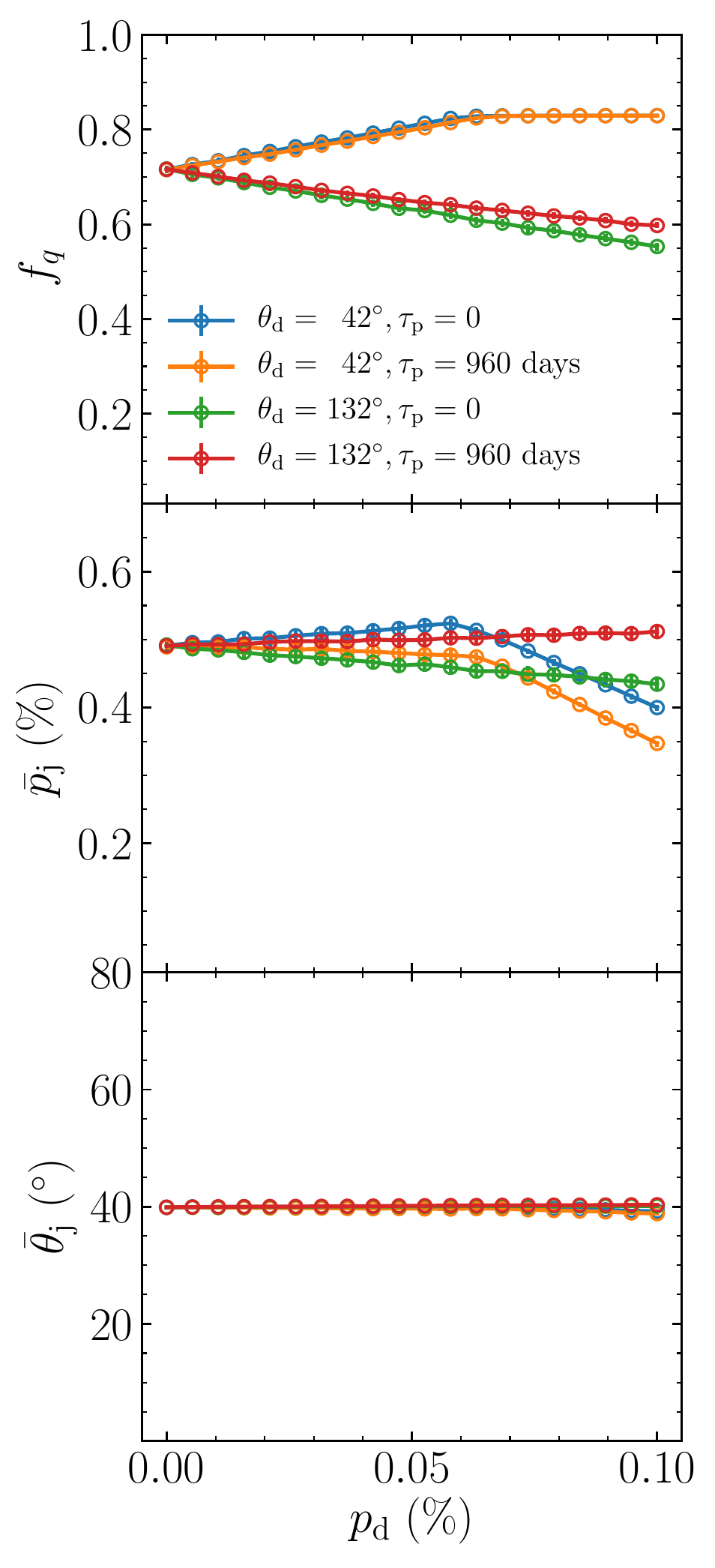}
\caption{The best estimated values of $f_{q}$, $\bar p_{\rm j}$, and $\bar \theta_{\rm j}$
for different input polarization degree of disk emission. Differently colored points 
represent different input polarization angle $\theta_{\rm d}$ of the disk emission and the
time delay $\tau_{\rm p}$ of the disk's polarized emission with respect to the unpolarized emission.}
\label{fig_pdisk}
\end{figure}

We by default neglect polarization of the disk emission in consideration 
of the nearly edge-on inclination ($i\sim10^\circ$) of the accretion disk. To see how this 
affects the obtained parameters, we input different polarization degree $p_{\rm d}$ of the disk emission
and show the results in Figure~\ref{fig_pdisk}. Previous studies demonstrated that 
the polarization angle of the disk emission should be either parallel or perpendicular to the 
the symmetry axis of the scattering region (\citealt{Goosmann2007,Li2009}), which is generally believed to align with 
the position angle of the jet. For 3C 273, the polarization angle of the disk emission 
is therefore either $42^\circ$ or $132^\circ$ (e.g., \citealt{Roeser1991}). Meanwhile,
if the polarization arises from scattering in equatorial torus, there might be a time delay between 
the polarized and the unpolarized disk emissions (\citealt{Gaskell2012, Rojas2020}). The size of the dust torus 
in 3C 273 inferred from near-infrared interferometry is about 960 days (\citealt{Kishimoto2011}).
Thus, we also show the results for cases of the time delay fixed to $\tau_{\rm p}=960$ days in Figure~\ref{fig_pdisk}.
Regarding the parameter $f_{\rm q}$ that is directly related to our flux decomposition, 
the best estimated value increases with $p_{\rm d}$ for cases of $\theta_{\rm d} = 42^\circ$
and approaches the upper limit of $f_{\rm q}\sim0.83$ when $p_{\rm d}>0.06\%$. 
For cases of $\theta_{\rm d} = 132^\circ$, the best estimated value of $f_{q}$ 
monotonously decreases with $p_{\rm d}$ and reaches $f_{q}\sim0.55$ for a moderate 
value of $p_{\rm d}=0.1\%$. The best estimate of $f_{q}$ is insensitive to $\tau_{\rm p}$.
As mentioned above, the disk emission with an inclination of $\sim10^\circ$ generally has a very low degree of
polarization because of symmetry. For example, the polarization degree of an optically 
thick disk is $\lesssim 0.05\%$ \citep[Section X]{Chandrasekhar1960}. We therefore expect 
a reasonable range of $f_{q}\sim(0.6-0.8)$.

\section{Discussions}
\subsection{The Radio Light Curve}
There are several programs that monitored 3C 273 at radio bands (e.g., \citealt{Aller1999,Lister2009, Richards2011}). We use the 15 GHz radio monitoring data 
from the Owens Valley Radio Observatory (\citealt{Richards2011}), which has the best sampling rate. According to the jet model developed by 
\cite{Turler1999}, the variability of jet emissions arises from a series of synchrotron outbursts. Each burst produces a light curve with a rapid rise and slow decay with 
time (see Figure 1 in \citealt{Turler1999}). The synchrotron turnover frequency decreases as the shock front propagates along the jet. 
The resulting light curve from this model has a larger variability at higher frequency (\citealt{Turler2000}). 
This means that the flux correlations between different frequencies may be not simply linear and the relation between the radio and optical 
fluxes in Equation (\ref{eqn_rj}) should be regarded as a zero-order approximation.

Meanwhile, at low radio frequency the outer jet and its terminal
hot spot may contribute mildly to the observed flux densities (e.g., \citealt{Bahcall1995,Conway1993, Perley2017}).
Therefore, it would be better to apply high-frequency radio light curves with sufficiently good 
sampling rate and long monitoring period in future once available.

\subsection{The Damped Random Walk Model}
We use the damped random walk model to describe the variability of the jet and disk emissions, which has been widely applied to 
AGN light curves (e.g., \citealt{Kelly2009, Zu2011, Li2013, Li2018, Pancoast2014}). However, there was evidence 
from the {\it Kepler} observations that AGN light curves no longer obey the damped random walk model on short timescales less than days
(\citealt{Kasliwal2015}). A more generic model would be the continuous-time autoregressive moving average (CARMA) model 
(\citealt{Kelly2014, Takata2018}), which can be regarded as a mixture of damped random walk models with different parameters.
It is possible to incorporate the CARMA model in our framework, but at the expense of massive computational overheads.
Indeed, different models mainly affect the short-timescale variability of the reconstructed light curves between measurement points.
The high sampling of our compiled light curve data helps to minimize this influence. Moreover, the convolution operations
in Equations (\ref{eqn_rj}) and (\ref{eqn_hb}) will also smooth the short-timescale variations to some extent. We therefore 
expect that the main results do not depend on the details of the adopted variability model.

\subsection{The Gaussian Transfer Functions}
\label{sec_tf}
For the sake of simplicity, we assume that both the transfer functions for the jet in Equation (\ref{eqn_rj}) and the broad-line region 
in Equation (\ref{eqn_hb}) are a single Gaussian. As such, all the covariance functions (see 
Appendix~\ref{sec_cov}) can be expressed analytically, which facilitates the calculations. There are two approaches for future improvements in practice.
First, using multiple Gaussians to model the real transfer functions (\citealt{Li2016}). This will retain the advantage that the 
covariance functions have analytical forms. Second, employing a physical jet model and broad-line model to directly calculate the transfer functions.
In particular for the broad-line region, there is a generic dynamical modeling method that works well for reverberation mapping data 
(\citealt{Pancoast2014, Li2013, Li2018}).
We expect that these two improvements could be beneficial to better fit the fine features in the light curves (in particular the H$\beta$ light curve),
however, the main results of the present calculations should be retained.

\begin{figure}
\centering 
\includegraphics[width=0.48\textwidth]{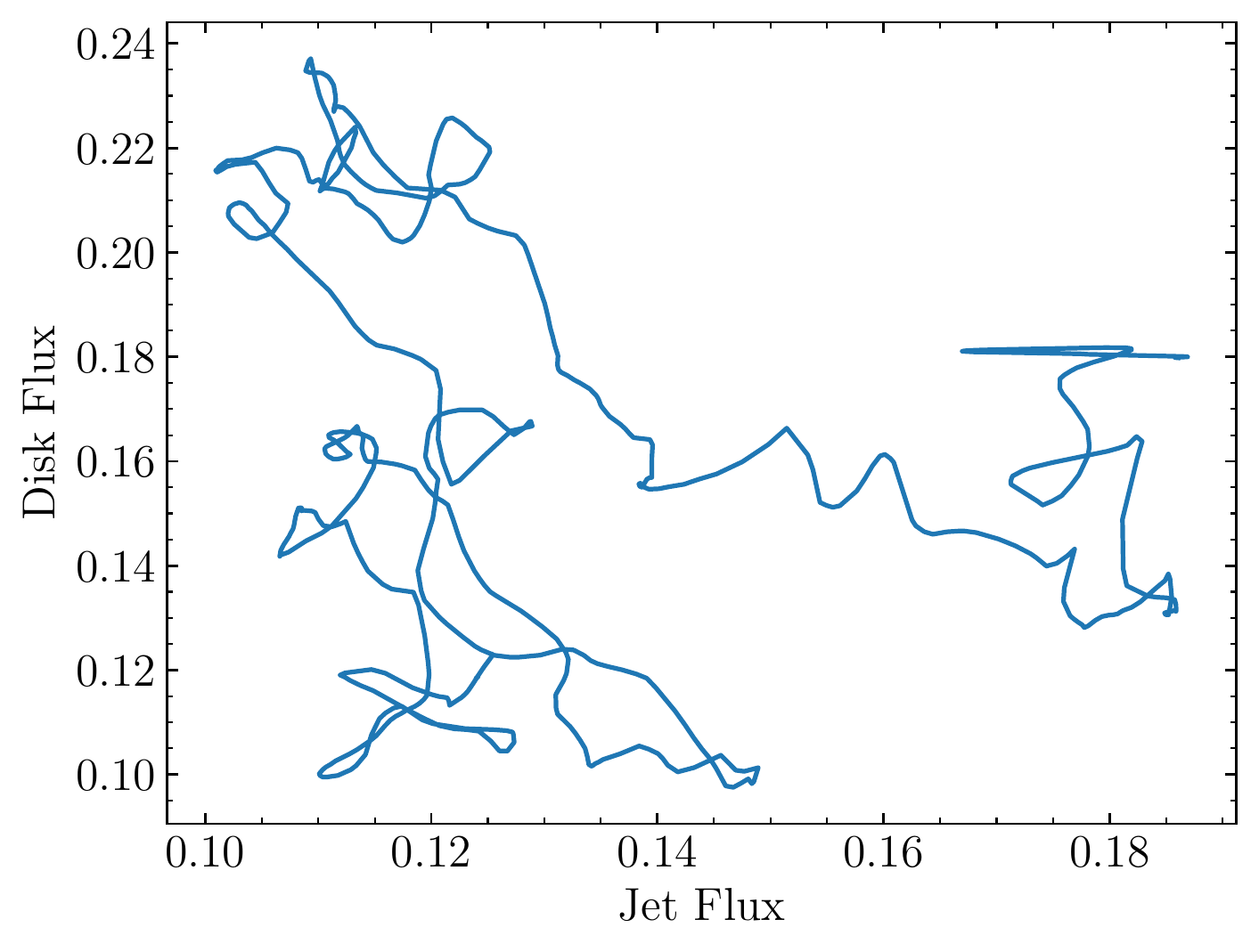}
\caption{The relation between the decomposed disk emission $C_{\rm d}(t)$ and jet emission $C_{\rm j}(t)$.}
\label{fig_dj}
\end{figure}

\subsection{The Accretion Disk-Jet Relation}
In Figure~\ref{fig_dj}, we plot the relation between the decomposed disk emission $C_{\rm d}(t)$ and jet emission $C_{\rm j}(t)$.
The evolutionary track is fully random, consistent with our assumption that $C_{\rm d}(t)$ and $C_{\rm j}(t)$ are independent. 
This reinforces that the variability of the disk and jet is stochastic and independent at short timescales, even though they maybe eventually connected 
over the lifetime of the black hole activity (far much longer than the temporal baseline of the present data and 
our results are therefore not affected). Meanwhile, such stochastic variability of individual objects will contribute to the intrinsic scattering of the so-called fundamental plane 
of black hole activity (e.g., \citealt{Merloni2003}).

\subsection{Implications for Reverberation Mapping Analysis}
The phenomenon that optical continuum emissions show long-term trends that do not have a corresponding echo in line emissions
is also incidentally detected in the past reverberation mapping campaigns (e.g., \citealt{Welsh1999, Denney2010, Li2013}). A conventional procedure that 
detrends the light curves of emission lines and continuum with low-order polynomials was usually used to remove the induced biases in 
cross-correlation analysis (\citealt{Denney2010,Peterson2014}). For the campaign of 3C 273 reported in \cite{Zhang2019}, 
if without detrending, the cross-correlation analysis on the light curves of the H$\beta$ line and the 5100~{\AA} continuum 
yields a time lag as large as 298 days (in observed frame) and a maximum correlation coefficient of $r_{\rm max}=0.7$. 
After detrending the light curve of the 5100~{\AA} continuum
with a linear polynomial, the maximum correlation coefficient increases to $r_{\rm max}=0.8$ and the time lag turns to be 170 days
(\citealt{Zhang2019}). The improvement on the maximum correlation coefficient illustrates that the detrending manipulation is 
necessary and worthwhile. Our work further reinforces such a detrending manipulation, however, the low-order polynomial should only be regarded as 
an approximation to the real trend. Indeed, we obtain a time lag of 194.9 days, larger by a factor of $\sim15$\% compared to the lag
determined with the linear detrending by \cite{Zhang2019}.
Like the case of 3C 273, multi-wavelength monitoring data are highly required to reveal 
the origin of non-echoed trends and therefore to conduct realistic, physical detrending.

On the other hand, the presence of non-echoed trends means that there exists a non-negligible component of continuum emissions not involved in photoionization 
of the broad-line region. As a result, this component of emissions needs to be excluded when positioning the objects in the size-luminosity 
scaling relation of broad-line regions. 
For 3C 273, the mean 5100~{\AA} flux is $19.46\times10^{-15}~{\rm erg~s^{-1}~cm^{-2}~\text{\AA}^{-1}}$ (\citealt{Zhang2019}),
corresponding a luminosity of $\log(\lambda L_{\lambda}/{\rm erg~s^{-1}}) = 45.86$ at 5100~{\AA} with a luminosity distance of 787 Mpc\footnote{We 
assume a standard $\Lambda$CDM cosmology with $H_0=67~{\rm km~s^{-1}~Mpc^{-1}}$, 
$\Omega_{\lambda}=0.68$, and $\Omega_{\rm M}=0.32$ (\citealt{Planck2014}).}.
Once excluding a mean fraction $(1-f_q)=0.284$ of the jet contribution, the realistic luminosity is changed to be 
$\log(\lambda L_{\lambda}/{\rm erg~s^{-1}}) = 45.72$, decreasing by about 0.14 dex. 
Accordingly, the dimensionless accretion rate $\mathscr{\dot M}=\dot M c^2/L_{\rm Edd}\propto (\lambda L_{\lambda})^{3/2}$ (e.g. \citealt{Du2014})
will decreases by about 0.2 dex, where $\dot M$ is the mass accretion rate and $L_{\rm Edd}$ is the Eddington luminosity.
This implies that for objects similar to 3C 273, it is important to 
correct for the jet or otherwise contaminations when applying the size-luminosity scaling relation.

\section{Conclusion}
3C 273 is a flat-spectrum radio quasar with both a blue bump and a beamed jet.
The recent reverberation mapping campaign reported by \cite{Zhang2019} showed that the optical continuum emissions display a non-echoed long-term trend 
compared to the emissions of the broad lines (such as H$\beta$ and \ion{Fe}{2}). 
In this work, we compile multi-wavelength monitoring data of 3C 273 from the {\it Swift} archive and other ground-based programs at optical and radio wavelengths 
(including the the reverberation mapping campaign).
The long-term trend of the {\it Swift} UV light curve is consistent with that of the H$\beta$ light curve but is clearly distinct from that of the optical light curves, 
exclusively indicating that there are two independent components of emissions at optical wavelength (see Section~\ref{sec_uv}). This is further reinforced by  
the complicated color variation behaviours and the low-level optical polarizations of 3C 273 (see Sections~\ref{sec_color} and \ref{sec_polar}).
Considering the coexistence of the comparably prominent jet and blue bump in 3C 273,  
these lines of observations pinpoint to two-fold origins of the optical emissions: one is the accretion disk itself
and the other is the jet.  

We developed an approach to decouple the optical emissions from the jet and accretion disk 
using the reverberation mapping data, 15 GHz radio monitoring data, and optical polarization data. 
We implicitly assume that the 15 GHz radio emission represents an blurred echo of the jet emission 
at optical wavelength with a time delay. The results show that the jet emissions can well explain the non-echoed long-term trend 
in the optical continuum (in terms of the H$\beta$ reverberation mapping). 
In consideration of the low inclination angle ($\sim10^\circ$) of the jet of 3C 273, we also simply assume that 
the disk emission has a negligible polarization. 
As a result, the jet quantitatively contributes 
a fraction of $\sim$10\% at the minimum and up to $\sim$40\% at the maximum to 
the total optical emissions. To our knowledge, this is the 
first time to interpret the conventional detrending procedure in reverberation mapping analysis with a physical process.
Our work generally supports the procedure in which low-order polynomials are adopted to detrend the light curves, however, we bear in mind the 
limited practicability of such low-order polynomials. To conduct realistic detrending, one generally needs multi-wavelength monitoring data, especially
UV data. Meanwhile, our work also implies that when applying the size-luminosity scaling relation for broad-line regions, 
one needs to carefully correct for the contaminations arising from non-echoed trends to optical luminosities.
This is particularly important for objects similar to 3C 273 with both prominent jet and disk emissions.

\acknowledgements
{
We thank the referee for constructive suggestions that
significantly improved the manuscript.
This research is supported in part by the National Key R\&D Program of China (2016YFA0400701), 
by the CAS Key Research Program (KJZDEW-M06), by the Key Research Program of
Frontier Sciences of the Chinese Academy of Sciences (QYZDJ-SSW-SLH007), and 
by grant No. NSFC-11833008, -11873048, -11991051, and -11991054 from the National Natural Science Foundation of China.
Y.R.L. acknowledges financial support from the National Natural Science Foundation of China through grant No. 11922304,
from the Strategic Priority Research Program of CAS through grant No. XDB23000000, and from the Youth Innovation Promotion Association CAS.  C.J. acknowledges financial support from 
the National Natural Science Foundation of China through grant No. 11873054.

This research has made use of data from the Steward Observatory spectropolarimetric monitoring project, which is 
supported by Fermi Guest Investigator grants NNX08AW56G, NNX09AU10G, NNX12AO93G, and NNX15AU81G.
This research has made use of up-to-date SMARTS optical/near-infrared light curves that are available at \url{www.astro.yale.edu/smarts/glast/home.php}.
This research has made use of data from the OVRO 40-m monitoring program, 
which is supported in part by NASA grants NNX08AW31G, NNX11A043G, and NNX14AQ89G and NSF grants AST-0808050 and AST-1109911.
}

\facility{SMARTS, Swift}
\software{\texttt{corner.py} (\citealt{Foreman2016}), \texttt{HEAsoft} (\citealt{Blackburn1995})}

\appendix
\section{Reduction for the Swift Data}
\label{sec_swift}
We used the HEASARC data archive to search for previous {\it Swift} observations of 3C 273 and download the data. 
We found 322 observations between 2004-12-13 and 2019-04-09, including 312 observations with exposures in both 
the X-ray Telescope (XRT) and the Ultraviolet/Optical Telescope (UVOT). We followed the standard threads\footnote{https://www.swift.ac.uk/analysis/uvot/}
and used \texttt{HEAsoft} (v6.25, \citealt{Blackburn1995}) to reduce the data. 
Firstly the {\sc xrtpipeline} script was used to reprocess the data with the latest calibration files. 
Then for each observation the {\sc uvotimsum} script was used to sum all exposures in every filter. 
Observations in the UVOT grism mode were excluded because we only wanted to use the six UVOT photometric 
bands (i.e. UVW2, UVM2, UVW1, $U$, $B$, $V$). Then a circular aperture with a radius of 5 arcsec was used to 
enclose the source region, while the background was extracted from a nearby circular region with a radius of 
20 arcsec without any point source. The {\sc uvotsource} script was used to determine the magnitude and flux 
in every filter. We also ran the small-scale sensitivity check to identify data points affected by the lower
throughput areas on the detector and discarded them. The final number of observations in every filter is 
listed in Table~\ref{tab_data} (not every observation had exposures in all six UVOT filters). Note that 3C 273 appears 
slightly extended in all six UVOT filters, and so the absolute source fluxes comprise both the AGN emission 
and part of the host galaxy star-light which we did not subtract, but the variability of the source flux should 
be attributed to the central AGN activity. 
We tabulated our reduced {\it Swift} UVOT fluxes of 3C 273 in Table~\ref{tab_uvot}, in which only a portion of the data is shown and the entire table 
is available in a machine-readable form online.

\section{Covariance Functions}
\label{sec_cov}
This appendix shows analytical expressions for the covariance functions in Section~\ref{sec_bayes}.
The covariance function between $C_{\rm t}(t)$  and $L_{\rm H\beta}(t)$ is 
\begin{equation}
S_{\rm tl}(\Delta t) = \langle C_{\rm t}(t_k) L_{\rm H\beta}(t_m)\rangle = \int \langle C_{\rm d}(t_k) C_{\rm d}(t_m)\rangle \Psi_{\rm H\beta}(t_m-\tau)d\tau,
\end{equation}
where $\Delta t = t_m - t_k$. With Equations (\ref{eqn_sdd}) and (\ref{eqn_tfd}), $S_{\rm tr}(\Delta t)$ has an analytical expression 
(see \citealt{Li2016} for a detailed derivation)
\begin{equation}
S_{\rm tl}(\Delta t) = \frac{1}{2}\sigma_{\rm d}^2 f_{\rm H\beta} e^{\omega_{\rm H\beta}^2/2T_{\rm d}^2}
\left\{ e^{-\Delta T/T_{\rm d}} {\rm erfc} \left[ -\frac{1}{\sqrt{2}}\left( \frac{\Delta T}{\omega_{\rm H\beta}} - \frac{\omega_{\rm H\beta}}{T_{\rm d}} \right) \right] 
       +e^{ \Delta T/T_{\rm d}} {\rm erfc} \left[  \frac{1}{\sqrt{2}}\left( \frac{\Delta T}{\omega_{\rm H\beta}} + \frac{\omega_{\rm H\beta}}{T_{\rm d}} \right) \right]\right\},
\label{eqn_stl}
\end{equation}
where $\Delta T = \Delta t - \tau_{\rm H\beta}$ and ${\rm erfc}(x)$ is the complementary error function. The auto-covariance function of $L_{\rm H\beta}(t)$ is given by
\begin{equation}
S_{\rm ll}(\Delta t) = \frac{1}{2}\sigma_{\rm d}^2 f_{\rm H\beta}^2 e^{\omega_{\rm H\beta}^2/T_{\rm d}^2}
\left\{ e^{-\Delta t/T_{\rm d}} {\rm erfc} \left[ - \frac{\Delta t}{2\omega_{\rm H\beta}} + \frac{\omega_{\rm H\beta}}{T_{\rm d}}  \right] 
       +e^{ \Delta t/T_{\rm d}} {\rm erfc} \left[   \frac{\Delta t}{2\omega_{\rm H\beta}} + \frac{\omega_{\rm H\beta}}{T_{\rm d}} \right]\right\}.
\label{eqn_sll}
\end{equation}
The covariance function $S_{\rm tr}(\Delta t)$ between $C_{\rm t} (t)$ and $R_{\rm j}(t)$ and the auto-covariance function $S_{\rm rr}(\Delta t)$ of $R_{\rm j}(t)$
can be expressed in similar forms by replacing the subscript ``$\rm H\beta$'' and ``d'' with ``j'' in Equations (\ref{eqn_stl}) and (\ref{eqn_sll}).

\begin{deluxetable*}{ccccccc}
\tablecolumns{7}
\tabletypesize{\footnotesize}
\tablecaption{{\it Swift} UVOT data of 3C 273. \label{tab_uvot}}
\tablehead{
\colhead{JD}   &
\colhead{UVW2}  &
\colhead{UVM2}  &
\colhead{UVW1}  &
\colhead{$U$}     &
\colhead{$B$}     &
\colhead{$V$}  \\\cline{2-7}
\colhead{($+$2,450,000)}  &
\multicolumn{6}{c}{($10^{-14}~\rm erg~s^{-1}~cm^{-2}~\text{\AA}^{-1}$)}
}
\startdata
3562.086 & \nodata & \nodata  &  \nodata &  $8.452\pm0.223$ & $4.982\pm0.095$  & $3.341\pm0.073$ \\
3562.152 & $19.011\pm0.548$ & $15.750\pm0.349$ & $13.219\pm0.477$ & \nodata & \nodata  &  \nodata \\
3689.602 & $21.013\pm0.607$ & $17.487\pm0.386$ & $14.281\pm0.518$ & $6.147\pm6.246$ & $4.378\pm0.092$ & $3.403\pm0.076$ \\
3692.082 & $21.157\pm0.627$ & \nodata          & \nodata          & \nodata         & \nodata         & $3.394\pm0.102$ \\
3721.746 & $20.954\pm0.608$ & $17.682\pm0.394$ & $14.491\pm0.528$ & $6.208\pm6.252$ & $4.871\pm0.104$ & $3.550\pm0.080$ \\
\nodata
\enddata
\tablecomments{This table is available in its entirety
in a machine-readable form in the online journal. Only a portion is shown here %
to illustrate its form and content.}
\end{deluxetable*}

\end{document}